%% file: article.tex
\documentclass[journal]{IEEEtran}
\usepackage{cite}
\ifCLASSINFOpdf
   \usepackage[pdftex]{graphicx}
   \graphicspath{{./}}
    \usepackage{epstopdf}
\else
\fi
\usepackage{amsmath}
\usepackage{amssymb}
  \usepackage[caption=false,font=footnotesize]{subfig}
\usepackage[per-mode=symbol]{siunitx}
\usepackage[utf8]{inputenc}
\usepackage{todonotes}
\graphicspath{{}}
\usepackage{bm}
\usepackage{mathtools}
\usepackage{setspace}
\usepackage{tikz}
\usepackage{threeparttable}
\usetikzlibrary{calc}
\usetikzlibrary{matrix,positioning}
\usepackage{pgfplots}
\usepgfplotslibrary{groupplots}
\pgfplotsset{compat=newest}
\usepackage[\ifnum\pdfoutput=1breaklinks\fi]{hyperref}
\usepackage{regexpatch}
\usepackage[acronym,nomain]{glossaries}
\glsdisablehyper
\makeglossaries
\loadglsentries{acronyms}

\usepackage{booktabs}
\usepackage{multirow}
\usepackage{url}

\input{commands}

\newsavebox{\myimage}

\begin{document}
\bstctlcite{IEEEexample:BSTcontrol}

\pgfplotscreateplotcyclelist{my cycle list}{%
	solid, black, every mark/.append style={solid}, mark=*\\%
	solid, red, every mark/.append style={solid}, mark=square*\\%
	solid, blue, every mark/.append style={solid}, mark=otimes*\\%
	solid, green, every mark/.append style={solid}, mark=triangle*\\%
	densely dashed, black, every mark/.append style={solid},mark=*\\%
	densely dashed, red, every mark/.append style={solid},mark=square*\\%
	densely dashed, blue, every mark/.append style={solid},mark=otimes*\\%
	densely dashed, green, every mark/.append style={solid},mark=triangle*\\%
	densely dotted, black, every mark/.append style={solid},mark=*\\%
	densely dotted, red, every mark/.append style={solid},mark=square*\\%
	densely dotted, blue, every mark/.append style={solid},mark=otimes*\\%
	densely dotted, green, every mark/.append style={solid},mark=triangle*\\%
}
\title{The World's First Real-Time Testbed for Massive MIMO: Design, Implementation, and Validation}

\author{
	\IEEEauthorblockN{Steffen Malkowsky$^{ 1}$, Joao Vieira$^{ 1}$, Liang Liu$^{ 1}$, Paul Harris$^{2}$, Karl Nieman$^{3}$, Nikhil Kundargi$^{3}$, Ian Wong$^{3}$, Fredrik Tufvesson$^{ 1}$, Viktor \"Owall$^{ 1}$, and Ove Edfors$^{ 1}$ } \\
	\IEEEauthorblockA{ $^{ 1}$ Dept. of Electrical and Information Technology, Lund University, Sweden\\
		$^{2}$ Communication Systems and Networks Group, University of Bristol, UK\\
		$^{3}$ National Instruments, Austin, Texas, USA\\ 
		firstname.lastname@\{eit.lth.se, bristol.ac.uk, ni.com\} } 
	}
\maketitle

\begin{abstract}
This paper sets up a framework for designing a massive \glsentryfull{mimo} testbed by investigating \gls{hw} and system-level requirements such as processing complexity, duplexing mode and frame structure.
Taking these into account, a generic system and processing partitioning is proposed which allows flexible scaling and processing distribution onto a multitude of physically separated devices.
Based on the given \gls{hw} constraints such as maximum number of links and maximum throughput for \glsentrylong{p2p} interconnections combined with processing capabilities, the framework allows to evaluate modular \gls{hw} components.
To verify our design approach, we present the LuMaMi (Lund University Massive MIMO) testbed which constitutes the first reconfigurable real-time \gls{hw} platform for prototyping massive MIMO.
Utilizing up to 100 \glsentrylong{bs} antennas and more than 50 \glsentrylongpl{fpga}, up to 12 \glsentrylongpl{ue} are served on the same time/frequency resource using an LTE-like \glsentrylong{ofdm} \glsentrylong{tdd}-based transmission scheme.
Proof-of-concept tests with this system show that massive MIMO can simultaneously serve a multitude of users in a static indoor and static outdoor environment utilizing the same time/frequency resource.
\end{abstract}
\glsresetall

\IEEEpeerreviewmaketitle

\section{Introduction}
\glsunset{mimo}
\IEEEPARstart{I}{n} \gls{mami} an unconventionally high number of \gls{bs} antennas (hundreds or even higher) is employed to serve \eg a factor of ten less \glspl{ue}.
Due to the excess number of \gls{bs} antennas, linear signal processing may be used to spatially focus energy with high precision, allowing to separate a multitude of \glspl{ue} in the spatial domain while using the same time/frequency resource~\cite{Marzetta2010}. 
\Gls{mami} theory promises a variety of gains, \eg increase in spectral and energy efficiencies as compared with single antenna and traditional MU-MIMO systems\cite{rusek2013,Rate}, thereby tackling the key challenges defined for 5G.
 
Although \gls{mami} is a promising theoretical concept, further development requires prototype systems for proof-of-concept and performance evaluation under real-world conditions to identify any further challenges in practice.
Because of its importance, both industry and academia are making efforts in building \gls{mami} testbeds, including the Argos testbed with 96-antennas\cite{Argos1}, Eurecom's 64-antenna \gls{lte} compatible testbed, Samsung's Full-Dimension (FD) \gls{mimo} testbed and Facebook's Project Aries. 
Nevertheless, publications systematically describing the design considerations and methodology of a \gls{mami} testbed are missing and real-time real-scenario performance evaluation of \gls{mami} systems using testbeds have not been reported yet. 
At Lund University, the first real-time \gls{mami} testbed, the \gls{lumami} testbed, showing successful \gls{mami} transmission on the \gls{ul}, was built\cite{Vieira2014a}.
Ever since, many testbeds have been constructed based on identical \gls{hw} utilizing the same generic design principle, \eg the \gls{mami} testbeds at the University of Bristol\cite{Bristol1}, Norwegian University of Science and Technology in Trondheim and University of Leuven in Belgium. 
The \gls{lumami} testbed provides a fully reconfigurable platform for testing \gls{mami} under real-life conditions.
To build a real-time \gls{mami} testbed many challenges have to be coped with.
For example, shuffling data from 100 or more antennas, processing large-scale matrices and synchronizing a huge number of physically separated devices.
All this has to be managed while still ensuring an overall reconfigurability of the system allowing experimental hardware and software solutions to be tested rapidly.

This paper discusses how implementation challenges are addressed by first evaluating high-level \gls{hw} and system requirements, and then setting up a generic framework to distribute the data shuffling and processing complexity in a \gls{mami} system based on the given \gls{hw} constraints for interconnection network and processing capabilities. 
Taking into account the framework and requirements, a suitable modular \gls{hw} platform is selected and evaluated.
Thereafter, a thorough description of the \gls{lumami} testbed is provided including system parameters, base-band processing features, synchronization scheme and other details.
The \gls{lumami} testbed constitutes a flexible platform that supports prototyping of up to 100-antenna \SI{20}{\mega\hertz} bandwidth \gls{mami}, simultaneously serving 12 \glspl{ue} in real-time using \gls{ofdm} modulation in \gls{tdd} transmission mode.
\Glspl{ber} and constellations for real-time \gls{ul} and \gls{dl} uncoded transmission in a static indoor and static outdoor scenario are presented.
Our first real-life proof-of-concept measurement campaigns show, that \gls{mami} is capable of serving up to 12 \glspl{ue} in the same time/frequency resource even for high user density per unit area.
The gathered results suggest a significant increase in spectral efficiency compared to traditional point-to-point \gls{mimo} systems.
By building the \gls{lumami} testbed we now have a tool which supports accelerated design of algorithms\cite{DBLP:journals/corr/VieiraREMLT16} and their validation based on real measurement data, with the additional benefit of real-world verification of digital base-band solutions.

Our main contributions can be summarized as follows:
\begin{itemize}
	\item We provide overall and thorough analysis for \gls{mami} systems, especially from a signal processing perspective, and identify design requirements as well as considerations on building up a \gls{mami} testbed.
	\item We propose signal processing breakdown and distribution strategy to master the tremendous computational complexity in a \gls{mami} system and introduce general hardware architecture for a \gls{mami} testbed.
	\item We present the world's first real-time 100-antenna \gls{mami} testbed, built upon \gls{sdr} technology.
	\item We validate the \gls{mami} concept and its spatial multiplexing capability in real-life scenarios (both indoor and outdoor) with over-the-air transmission and real-time processing. 
\end{itemize}

\section{Massive MIMO Basics}
\label{sec:MaMi_Theory}
\begin{figure}[!t]
	\centering
	\includegraphics[width=\linewidth]{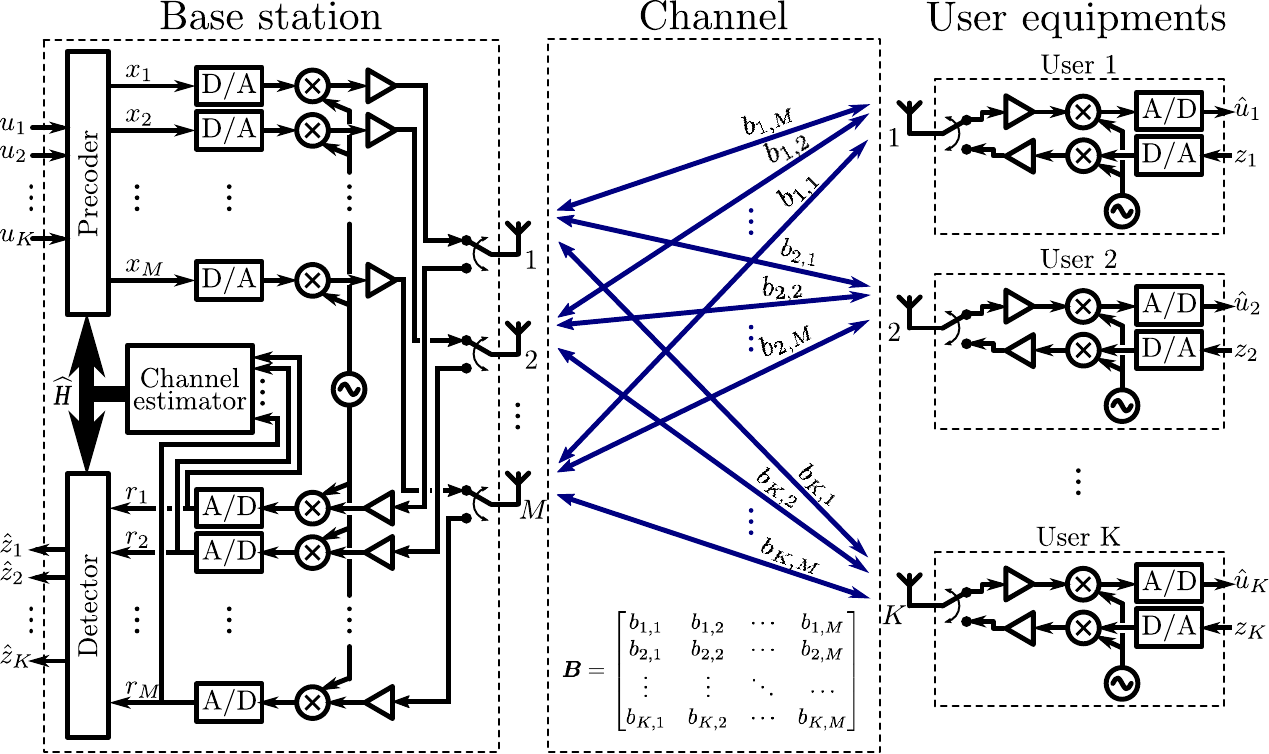}
	\caption{A \gls{mami} system model. Each antenna at the BS (left side) transmits a linear combination of $K$ user-intended data symbols ${u_k}_{k=1}^K$. After propagation through the DL wireless channel $\bm{B}$, each user antenna receives a linear combination of the signals transmitted by the M BS antennas. Finally, each of the K users, say user k, produces an estimate of its own intended data symbol, \ie $u_k$. Similar operation is employed for UL data transmission. Here, reciprocity for the propagation channel is assumed, \ie $\bm{B}=\bm{B}^\mathsf{T}$.}
	\label{fig:system_model}
\end{figure}
In this section, the basic key detection and precoding algorithms utilized in \gls{mami} are presented.
Implementation specific details required to apply these algorithms, such as \gls{csi} estimation, are discussed in Sec.~\ref{sec:lumami_testbed_implementation}.
A simplified model of a \gls{mami} \gls{bs} using $M$ antennas while simultaneously serving $K$ single antenna \glspl{ue} in \gls{tdd} operation in a propagation channel $\bm{B}$ is shown in \figurename~\ref{fig:system_model}.
To simplify notation, this discussion assumes a base-band equivalent channel and expressions are given per subcarrier, with subcarrier indexing suppressed throughout.

\begin{table}[t]
	\centering
	\renewcommand{\arraystretch}{1.3}
	\setlength{\abovecaptionskip}{0pt}
	\footnotesize
	\caption{Linear Precoding/Detection Matrices}
	\scalebox{0.95}{
	\noindent\begin{tabular}{cccc}
		\toprule
		& \textbf{MRT/MRC} & \textbf{ZF} & \textbf{RZF} \\
		\midrule
		DL & $\bm{C}\bm{G}^*$ & $\bm{C}\bm{G}^*(\bm{G}^\mathsf{H}\bm{G})^{\mathsf{-T}}$ & $\bm{C}\bm{G}^*(\bm{G}^\mathsf{H}\bm{G} + \beta_{\mathrm{reg}_\mathrm{pre}}\bm{I}_{K})^{\mathsf{-T}}$\\
		UL & $\bm{G}^\mathsf{H}$ & $(\bm{G}^\mathsf{H}\bm{G})^{-1}\bm{G}^\mathsf{H}$ & $(\bm{G}^\mathsf{H}\bm{G} + \beta_{\mathrm{reg}_\mathrm{dec}}\mathbf{I}_{K})^{-1}\bm{G}^\mathsf{H}$ \\
		\bottomrule
	\end{tabular}}
	\label{tab:Linear_Det_Pre_Schemes}
\end{table}

\subsection{Up-link}
The \gls{ul} power levels used by the $K$ \glspl{ue} during transmission build the $K \times K$ diagonal matrix $\bm{P}_\mathrm{ul}$. By collecting the transmitted \gls{ue} symbols in a vector $\bm{z} \triangleq (z_1,\ldots,z_K)^\mathsf{T}$, the received signals $\bm{r} \triangleq (r_1,\ldots,r_M)^\mathsf{T}$ at the \gls{bs} are described as
\begin{align}\label{eq:ul_model}
    \bm{r} = \bm{G}\sqrt{\bm{P}_\mathrm{ul}}\bm{z} + \bm{w},
\end{align}
where $\bm{G}$ is the $M\times K$ \gls{ul} channel matrix\footnote{$\bm{G}$ is the up-link radio channel capturing both, the propagation channel $\bm{B}^\mathsf{T}$ and the up-link hardware transfer functions.}, $\sqrt{\bm{P}_\mathrm{ul}}$ an elementwise square-root, and $\bm{w} \sim  \mathcal{CN}(0,\mathbf{I}_M)$ is \gls{iid} circularly-symmetric zero-mean complex Gaussian noise.  
The estimated user symbols $\widehat{\bm{z}} \triangleq (\hat{z_1},\ldots,\hat{z_K})^\mathsf{T}$ from the $K$ \glspl{ue} are obtained by linear filtering of the received vector $\bm{r}$ as
\begin{align}
\label{eq:Detector}
    \widehat{\bm{z}} = f_\text{eq}(\bm{G})\bm{r},
\end{align}
where  $f_\text{eq}( \cdot )$ constructs an appropriate equalization matrix.

\subsection{Down-link}
On the \gls{dl}, each \gls{ue} receives its corresponding symbol $\hat{u}_k$ which are collected in a vector $\widehat{\bm{u}} \triangleq  (\hat{u}_1, \ldots ,\hat{u}_K)^\mathsf{T}$, representing the symbols received by all \glspl{ue}. With this notation, the received signal becomes
\begin{align}
\label{eq:DLchannel}
\widehat{\bm{u}} = \bm{H}\bm{x} + \bm{w}'
\end{align}
where the $K \times M$ matrix $\bm{H}$ is the \gls{dl} radio channel\footnote{$\bm{H}$ is the down-link radio channel capturing both, the propagation channel $\bm{B}$ and the down-link hardware transfer functions.},  $\bm{w}' \sim  \mathcal{CN}(0,\mathbf{I}_K)$ is an \gls{iid} circularly-symmetric zero-mean complex Gaussian receive noise vector with covariance matrix $\mathbf{I}_K$, and ${\bm{x} \triangleq  (x_1, \ldots , x_M)^\mathsf{T}}$ is the transmit vector. 

As explicit \gls{dl} channel estimation is very resource consuming, it is not considered practical in a \gls{mami} setup\cite{Marzetta2010}.
Taking into account that the propagation channel $\bm{B}$ is generally agreed on to be reciprocal\cite{DBLP:journals/corr/VieiraREMLT16}, the estimated \gls{ul} channel matrix $\bm{G}$ can be utilized to transmit on the \gls{dl}.
However, differences due to analog circuitry in the \gls{ul} and \gls{dl} channels, $\bm{G}$ and $\bm{H}$, need to be compensated.
Thus, a possible construction for $\bm{x}$ is of the form
\begin{align}
\label{eq:SteffensAss}
\bm{x} =   f_\text{cal}(  f_\text{pre}(\bm{G}))\bm{u},
\end{align}
where $\bm{u} \triangleq  (u_1, \ldots ,u_K)^\mathsf{T}$ is a vector containing the symbols intended for the $K$ \glspl{ue}, $f_\text{pre}( \cdot )$ is some precoding function, and $f_\text{cal}( \cdot )$ is a reciprocity calibration function to be discussed next.

\subsection{Reciprocity Calibration}
\label{subsec:rec_cal}

In most practical systems, the \gls{ul} and \gls{dl} channels are not reciprocal, i.e. $\bm{G} \neq \bm{H}^T$. This is easily seen by factorizing $\bm{G}$ and $\bm{H}$ as
\begin{equation}
\label{eq:Gfact}
\bm{G} = \bm{R}_{\rm B}  \bm{B}^\mathsf{T} \bm{T}_{\rm U}, \; \; \; {\rm and} \; \; \; \bm{H} = \bm{R}_{\rm U}  \bm{B} \bm{T}_{\rm B},
\end{equation}
where the two $M\times M$ and $K\times K$ diagonal matrices $\bm{R}_{\rm B}$ and $\bm{R}_{\rm U}$ model the non-reciprocal hardware responses of \gls{bs} and \gls{ue} \glspl{rx}, respectively, and the two $M\times M$ and $K\times K$ diagonal matrices $\bm{T}_{\rm B}$ and $\bm{T}_{\rm U}$ similarly model hardware responses of their \glspl{tx}. Thus, in order to construct a precoder based on the \gls{ul} channel estimates, the non-reciprocal components of the channel have to be calibrated. Previous calibration work showed that this is possible by using 
\begin{equation}
\label{eq:Calibration}
\bm{C} f_\text{pre}(\bm{G}) = f_\text{cal}(  f_\text{pre}(\bm{G})),
\end{equation}
where $\bm{C}=\bm{R}_{\rm B} \bm{T}^{-1}_{\rm B}$ is the, so-called, calibration matrix which can be estimated internally at the \gls{bs}\cite{DBLP:journals/corr/VieiraREMLT16}. Such calibration is sufficient to cancel inter-user interference stemming from non-reciprocity\cite{6760595}.

\subsection{Linear Detection \& Precoding Schemes}
Table \ref{tab:Linear_Det_Pre_Schemes} shows a selection of weighting matrices used in linear precoding and detection schemes, with non-reciprocity compensation included in the form of the $M \times M$ diagonal matrix $\bm{C}$ as defined above.
The \gls{mrt} precoder and the \gls{mrc} decoder maximize array gain without active suppression of interference among the \glspl{ue} \cite{Marzetta2010}.
The \gls{zf} precoder and \gls{zf} combiner employ the pseudo-inverse, which provides inter-user interference suppression with the penalty of lowering the achievable array gain.
A scheme that allows trade-off between array gain and interference suppression is the \gls{rzf} precoder and \gls{rzf} combiner.
This is achieved by properly selecting the regularization constants $\beta_{\mathrm{reg}_\mathrm{pre}}$ and $\beta_{\mathrm{reg}_\mathrm{dec}}$.
If $\beta_{\mathrm{reg}_\mathrm{pre}}$ and $\beta_{\mathrm{reg}_\mathrm{dec}}$ are selected to minimize \gls{mse} $E{\|\bm{u} - \tfrac{1}{\sqrt{\rho}}\hat{\bm{u}}\|^2}$, where $\rho$ is a scaling constant, we obtain the \gls{mmse} precoder/detector \cite{Bjoernson2014a}.

\section{System Design Aspects}
\label{sec:System_Design_Aspect}

Having discussed the \gls{mami} basics, we move on to system design aspects.
These include modulation scheme, frame structure and hardware requirements.

\subsection{Modulation Scheme} 
While many different modulation schemes can be used with \gls{mami}, this paper focuses on \gls{ofdm}, employed in many modern wireless communication systems.
Properly designed \gls{ofdm} renders frequency-flat narrowband subcarriers, facilitating the single channel equalization strategy used here.

For ease of comparison and simplicity, \gls{lte}-like \gls{ofdm} parameters, as shown in Table~\ref{table:SystemParam}, are used throughout this discussion.
The more common parameters with \gls{lte}, the easier it is to evaluate how \gls{mami} as an add-on would influence current cellular systems.
\begin{table}[t]
	\footnotesize
	\caption{High-level system parameters}
	\noindent\begin{tabular*}{\columnwidth}{@{\extracolsep{\stretch{1}}}*{7}{l}@{}}
		\toprule
		\textbf{Parameter} & \textbf{Variable} & \textbf{Value} \\
		\midrule
		Bandwidth & $W$ & 20\,MHz \\
		Sampling Rate & $F_\text{s}$ & 30.72\,MS/s \\
		FFT Size & $N_\text{FFT}$ & 2048 \\
		\# Used subcarriers & $N_\text{used}$  & 1200 \\
		Cyclic prefix & $N_\text{cp}$  & 144 samples \\
		OFDM symbol length & $t_\text{\tiny{OFDM}}$ & \SI{71.4}{\micro\second}\\
		\bottomrule
	\end{tabular*}
	\label{table:SystemParam}
\end{table}

\subsection{TDD versus FDD}
Current cellular systems either operate in \gls{fdd} or \gls{tdd} mode.
\gls{fdd} is, however, considered impractical for \gls{mami} due to excessive resources needed for \gls{dl} pilots and \gls{csi} feedback. \gls{tdd} operation relying on reciprocity only requires orthogonal pilots in the \gls{ul} from the $K$ \glspl{ue}, making it the feasible choice \cite{7402270}. For this reason, we focus entirely on \gls{tdd} below.

\subsection{Reciprocity}
To allow operation in \gls{tdd} mode, differences in the \gls{tx} and \gls{rx} transfer functions on both, the \gls{bs} and \glspl{ue} have to be calibrated as discussed in Sec.~\ref{subsec:rec_cal}.
Drifts over time are mainly caused by \gls{hw} temperature and voltage changes, and thus, the calibration interval depends on the operating environment of the \gls{bs}.

\subsection{Frame Structure}
The frame structure defines among other things, the pilot rate which determines how well channel variations can be tracked and, indirectly, the largest supported UE speed.

\subsubsection{Mobility}
The maximum supportable mobility, \eg the maximum speed of the \glspl{ue} is defined by the \gls{ul} pilot transmission interval.
In order to determine this constraint, a 2D wide-sense stationary channel with uncorrelated isotropic scattering is assumed.
For the contributions from the different \gls{bs} antennas to add up coherently high channel correlation is required and, as an approximation to formulate the final requirement, a correlation of $0.9$ was used to ensure sufficient channel coherency.
Further discussions on such modeling assumption are found in\cite{molisch2010wireless}.
Although these assumptions may not be completely valid for \gls{mami} channels, they allow an initial evaluation based on a maximum supported Doppler frequency, $\nu_\text{max}$, by solving 
\begin{align}
J_\text{0}(2\pi\nu_\text{max}T_\text{p}) = 0.9,\label{eq:bessel}
\end{align}
for $\nu_\text{max}$, where $J_\text{0}(\cdotp)$ is the zeroth-order Bessel function of the first kind, stemming from a standard Jakes' fading assumption, and $T_\text{p}$ the distance between pilots in time.
Hence, the maximum supportable speed of any \gls{ue} may be evaluated using
\begin{align}
\text{v}_\text{max}= \frac{c\nu_\text{max}}{f_\text{c}},\label{eq:Mobility}
\end{align}
once a specific frame structure is provided.
In (\ref{eq:Mobility}) $\text{v}_\text{max}$ is the maximum supported speed of a \gls{ue}, $c$ the speed of light and $f_c$ the chosen carrier frequency.
\subsubsection{Processing latency}
The frame structure has to be designed for the highest speed of \glspl{ue} to be supported which requires a high pilot rate for high mobility scenarios. 
Within two consecutive \gls{ul} pilot symbols, all \gls{ul} data, \gls{dl} data and guard symbols have to be accommodated	 which in turn decreases the available time between UL pilot reception and DL transmission. 
In a high mobility scenario this poses tight latency requirements for \gls{tdd} transmission as \gls{csi} has to be estimated in order to produce the precoding matrix to beamform the \gls{dl} data.

To formulate the \gls{tdd} precoder turnaround time, $\Delta$, all \gls{hw} units introducing a delay must be taken into account.
This includes the analog front-end delays for the \gls{tx} $\Delta^\text{rf,\tiny{TX}}$ and \gls{rx} $\Delta^\text{rf,\tiny{RX}}$, the processing latency for \gls{ofdm} modulation/demodulation (including \gls{cp} and guard band operation) $\Delta^\text{\tiny{OFDM}}$, the time for processing \gls{ul} pilots to estimate \gls{csi} $\Delta^\text{\tiny{CSI}}$, and the processing latency for precoding $\Delta^\text{precode}$ including reciprocity compensation.
Additional sources of latency include overhead in data routing, packing, and unpacking, \ie $\Delta^\text{rout}$ such that the overall \gls{tdd} precoder turnaround time may be formulated as
\begin{equation}
\Delta = \Delta^\text{rf,\tiny{TX}}+\Delta^\text{rf,\tiny{RX}}+\Delta^\text{\tiny{OFDM}}+\Delta^\text{\tiny{CSI}} + \Delta^\text{precode} + \Delta^\text{rout}.
\label{eq:latency}
\end{equation}
Depending on the specific arrangement of the \gls{ofdm} symbols and the pilot repetition pattern in the frame structure, base-band processing solutions, especially $\Delta^\text{\tiny{CSI}}$ and $\Delta^\text{precode}$, have to be optimized to not violate the given constraint, \ie $\Delta$. 

\subsubsection{Pilot pattern}
In general, to acquire \gls{csi} at the \gls{bs}, the $K$ \glspl{ue} transmit orthogonal pilots on the \gls{ul}.
Different approaches are, \eg distributed pilots over orthogonal subcarriers\cite{deliverable3_2} or sending orthogonal pilot sequences over multiple subcarriers\cite{6783987,Pilot1,Pilot2} but also semi-blind and blind techniques have been proposed\cite{7339665}.

\begin{figure}[!t]
	\centering
	\includegraphics[width=0.9\columnwidth]{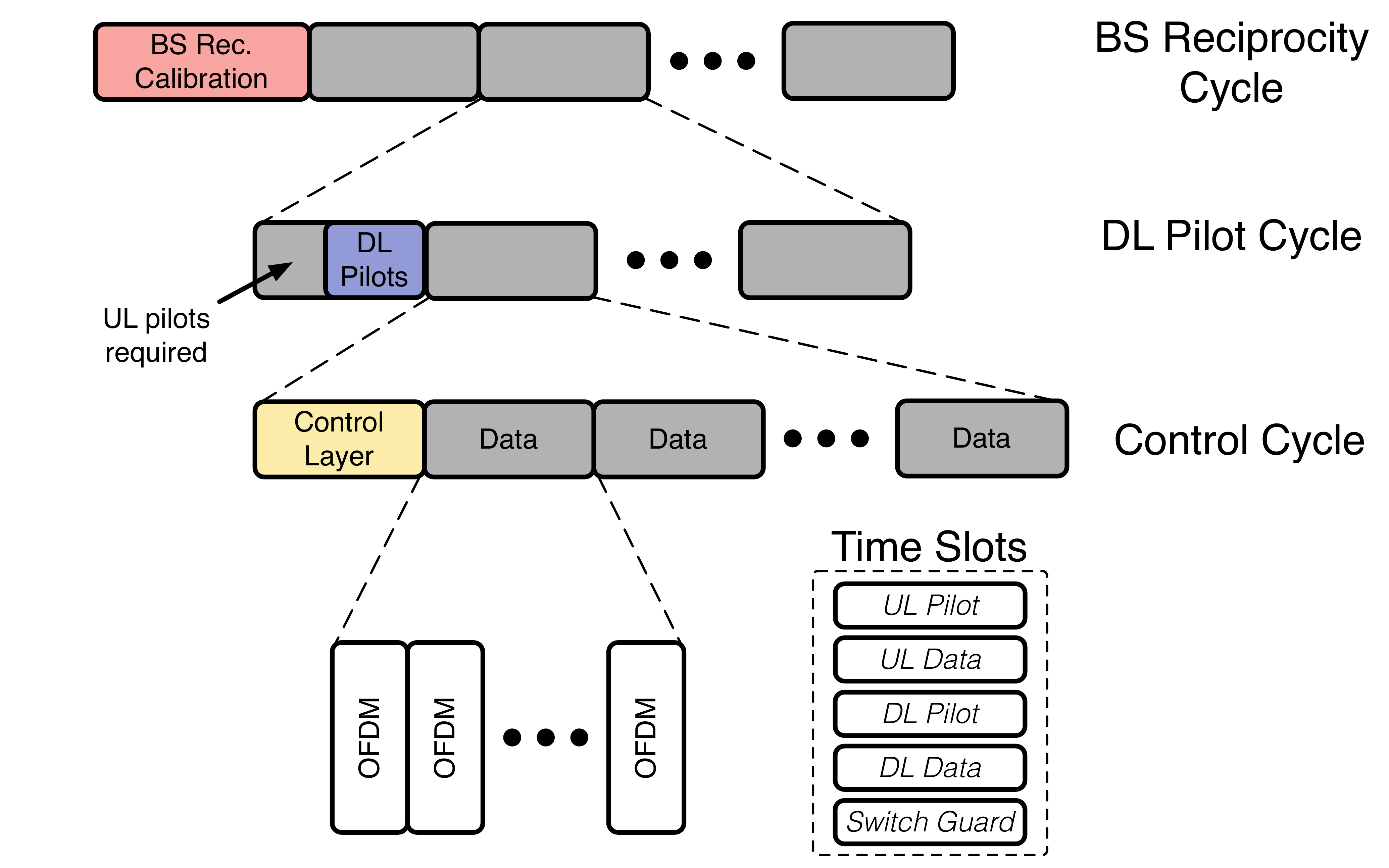}
	\caption{Generic frame structure of a \gls{lte} like \gls{tdd}-based \gls{mami} system. Within one BS reciprocity cycle the BS operates using the same reciprocity calibration coefficients. A certain number of DL pilot cycles are integrated as UEs suffer from faster changing environments. Each control cycle contains a control layer to perform, for example over-the-air synchronization and within these the data transmission slots are encapsulated.}
	\label{fig:frame_structure_generic}
\end{figure}

\figurename~\ref{fig:frame_structure_generic} shows a generic frame structure capturing the aforementioned aspects in a hierarchical manner assuming all \glspl{ue} transmit their pilots within one dedicated pilot symbol.
At the beginning of each \gls{bs} reciprocity cycle, reciprocity calibration at the \gls{bs} is performed and within these a certain number of \gls{dl} pilot cycles are encapsulated where precoded \gls{dl} pilot symbols are transmitted.
The length of the BS reciprocity cycle is determined by the stability of the transceiver chains in the \gls{bs}.
As the reciprocity calibration at the \gls{bs} side only compensates for \gls{bs} transceivers, \gls{dl} pilots are necessary to compensate for transceiver differences at the \gls{ue} side.
Their frequency depends on the stability at the \gls{ue} side and can be considered significantly smaller than for the \gls{bs} as \glspl{ue} are subject to faster changes in their operational environment, \eg thermal differences when having the \gls{ue} in a pocket or using it indoors or outdoors.
To be able to send precoded pilots on the \gls{dl}, transmission of \gls{ul} pilots is required beforehand.
Several control cycles are embedded inside each \gls{dl} pilot cycle carrying a certain number of data time slots.
Time slots contain five different \gls{ofdm} symbol types for physical layer implementation.
These are \emph{(i)} UL Pilot where the \glspl{ue} transmit orthogonal pilots to the \gls{bs}, \emph{(ii)} UL Data where all \glspl{ue} simultaneously send data to the \gls{bs}, \emph{(iii)} DL Pilot where the \gls{bs} sends precoded pilots to all \glspl{ue}, \emph{(iv)} DL Data where the \gls{bs} transmits data to all \glspl{ue} and \emph{(v)} Switch Guard, which idles the RF chains to allow switching from RX to TX or vice versa.

\subsection{Hardware Requirements}
To illustrate the required \gls{hw} capabilities for the testbed, the values from Table~\ref{table:SystemParam} are used to estimate the \SI{}{Gops\per\second}
\footnote{\SI{}{Gops\per\second} is used here, but these can be seen as \SI{}{GMACs\per\second}, \ie the number of multiply-accumulate operations, as almost all operations involve matrix-matrix and matrix-vector calculations.} 
and the data shuffling on a per \gls{ofdm} symbol basis for the general case and a specific case assuming $M=100$ and $K=12$.

\subsubsection{Processing Capabilites}
\label{subsubsec:Processing_Capabilities}
Table~\ref{tab:hw_requirements} summarizes the overall number of real-valued arithmetic operations.
For the processing estimates, it is assumed that each complex multiplication requires four real multiplications.
Close to the antennas, $M$ \glspl{fft} or \glspl{ifft} are needed equating to \SI{126}{Gops\per s}.
Data precoding and detection as well as reciprocity compensation require large matrix and vector multiplications, for instance, an $M\times K$ matrix with a $K\times 1$ vector leading to up to \SI{80}{Gops\per s}.

Finally, when using ZF, the pseudo-inverse matrix is required which includes the calculation of the Gram matrix requiring $MK^2$ multiplications with the $K\times K$ matrix inversion adding another $K^3$ in complexity assuming a Neumann-Series approximation\cite{Prabhu2013} or a QR decomposition. The last multiplication of the inverse with the Hermitian of the channel matrix $\bm{H}$ needs another $MK^2$ multiplications which combined with a requirement of finishing within two \gls{ofdm} symbols leads to approximately \SI{1}{Tops\per\second} for the overall pseudo-inverse calculation.
\begin{table}[t]
	\centering
	\footnotesize
	\caption{Processing Requirements in a MaMi system}
	\noindent\begin{tabular}{llll}
		\toprule
		\textbf{Function} & \textbf{General} & \textbf{Specific} \\
		\midrule
		& \SI{}{Gops\per s} & \SI{}{Gops\per s}\\
		FFT/IFFT & $4 M \log_2(N_\text{FFT})N_\text{FFT}/t_\text{\tiny{OFDM}}$ & \num{126} \\
		Detection & $4 M K N_\text{used}/t_\text{\tiny{OFDM}}$ & \num{80} \\
		Precoding & $4 M K N_\text{used}/t_\text{\tiny{OFDM}}$ & \num{80} \\
		Recip. Cal. & $4 M K N_\text{used}/t_\text{\tiny{OFDM}}$ &\num{80} \\
		Pseudo-inv. & $4 N_\text{used}\left(2 MK^2 + K^3\right)/\left(2t_\text{\tiny{OFDM}}\right)$ & \num{1080}\\
		\bottomrule
	\end{tabular}
	\label{tab:hw_requirements}
\end{table}

\subsubsection{Data Shuffling Capabilities}
Table~\ref{tab:data_shuffling_requirements} summarizes required interconnect bandwidth and number of links.
Communication paths to each antenna transfer at the sampling rate of %
$F_\text{s}=\SI{30.72}{\mega S\per\second}$ which is decreased to the subcarrier rate %
$F_\text{sub}=\SI{16.8}{\mega B\per\second}$ by performing \gls{ofdm} processing $\left(F_\text{s}\cdot N_\text{used}/(N_\text{FFT}+N_\text{cp})\right)$.
Considering $M$ antennas, the overall subcarrier data rate is $M\cdot w\cdot\SI{16.8}{\mega B\per\second}$, with $w$ being the combined wordlength for the in-phase and quadrature components in bytes. %
The information rate in an \gls{ofdm} symbol carrying data is $K\cdot\SI{16.8}{\mega B\per\second}$ assuming $8$ bit per sample, \ie $256-$QAM as highest modulation.
Assuming separate links between centralized processing and the antenna units on \gls{ul} and \gls{dl}, $2M$ \gls{p2p} links\footnote{In this discussion, each interconnection transferring data between physically separated devices is denoted a \glsentryfull{p2p} link.} are needed between the antennas and the centralized \gls{mimo} processing.
\begin{table}[t]
	\centering
	\footnotesize
	\caption{Data Shuffling Requirements in a MaMi system}
	\noindent\begin{tabular}{lll}
		\toprule
		\textbf{Purpose} & \textbf{General} & \textbf{Specific} \\
		\midrule
		& \# & \# \\
		Links to cent. proc & $2M$ & 200\\
		\midrule
		& \SI{}{MB\per s} & \SI{}{MB\per s}\\
		Antenna Rate & $w_\text{ant}MF_\text{s}$ & $w_\text{ant}$ 3,072\\
		Subcarrier Rate & $wMF_\text{sub}$ & $w$ 1,680\\
		Information rate & $K\cdot F_\text{sub}$ & $201.6$ \\
		\bottomrule
	\end{tabular}
	\label{tab:data_shuffling_requirements}
\end{table}

\subsubsection{Reconfigurability}
The testbed has to be reconfigurable and scalable, to support different system parameters, different processing algorithms and adaptive processing.
It is also crucial to have the possibility to integrate in-house developed \gls{hw} designs for validation and performance comparison of algorithms.
Variable center frequencies, run-time adjustable \gls{rx} and \gls{tx} gains as well as configurable sampling rates are highly desirable to be able to adapt to other parameters than the ones presented in Table~\ref{table:SystemParam}.

\section{Generic Hardware and Processing Partitioning}
\label{sec:Generinc_Partitioning}
In this section a generic \gls{hw} and processing partitioning is presented
to explore the parallelism in \gls{mami}, which needs consideration of processing together with data transfer requirements (throughput, latency, \# of \gls{p2p} links), and at the same time provides scalability.

\subsection{Hierarchical Overview}
To be able to build a \gls{mami} testbed with modular \gls{hw} components, a hierarchical distribution as shown in \figurename~\ref{fig:Hierarchical_Overview} is proposed.
The main blocks are detailed as follows:
\begin{figure}
	\centering
	\includegraphics[width=0.85\columnwidth]{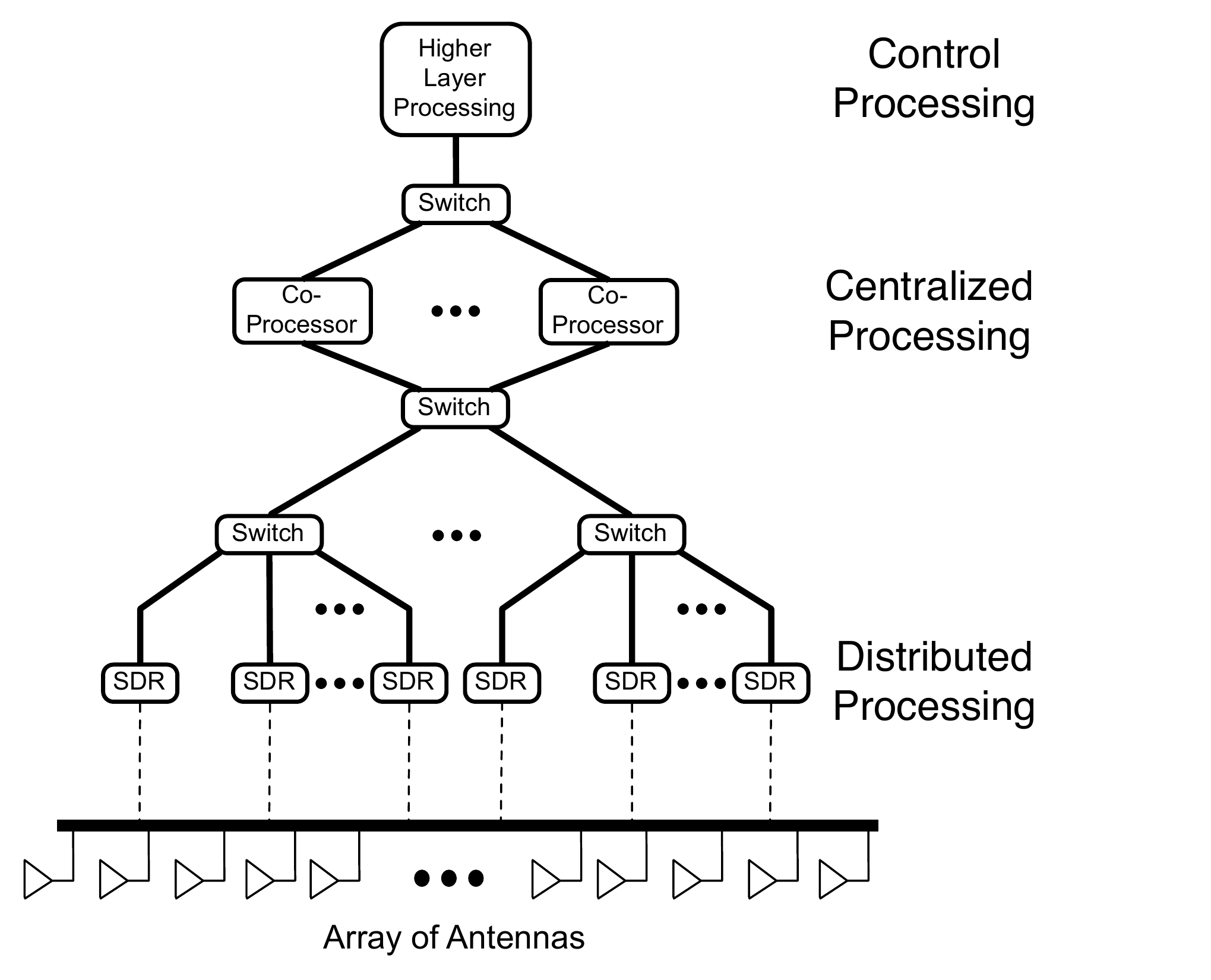}
	\caption{Hierarchical overview of a \gls{mami} \gls{bs} built from modular \gls{hw} components.}
	\label{fig:Hierarchical_Overview}
\end{figure}

\subsubsection{SDR} \glspl{sdr} provide the interface between the digital and \gls{rf} domain as well as local processing capabilities.

\subsubsection{Switches} Switches aggregate/disaggregate data between different parts of the system, \eg between \glspl{sdr} and the co-processors.

\subsubsection{Co-processing modules}
Co-processing modules provide a centralized node to perform \gls{mimo} processing.

\subsubsection{Higher Layer Processing} 
Higher layer processing controls the system, configures the radios, and provides run-time status metrics of the system.

\subsection{Processing and Data Distribution}
For proper base-band processing partitioning, throughput constraints of \gls{hw} components have to be taken into account.
Assuming each \gls{sdr} supports $n_\text{ant}$ antennas, the required number of \glspl{sdr} becomes $\lceil M/n_\text{ant}\rceil$ for an $M$-antenna system.

\subsubsection{Subsystems}
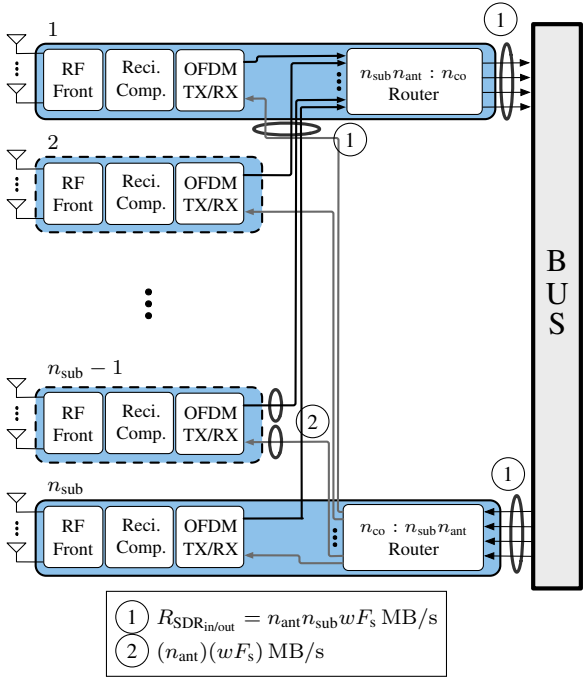
\begin{figure}
	\centering
	\hspace*{0pt}\input{subsystem.tikz}
	\caption{A subsystem consisting of $n_\text{sub}$ \glspl{sdr} where the two outer \glspl{sdr} implement an antenna combiner / BW splitter and an antenna splitter / BW combiner, both implemented using high-speed \glspl{fpga} routers. Inter-SDR and SDR to central processor connections utilize a bus for transferring the samples.}
	\label{fig:subsystem}
\end{figure}

As shown in \figurename~\ref{fig:subsystem}, RF-Front End, \gls{ofdm} processing and reciprocity compensation are performed on a per-antenna basis using the \glspl{sdr}.
This distributes a large fraction of the overall processing and reduces the data rate before transferring the acquired samples over the bus.
Still, the number of direct devices on a bus is limited, and thus, setting up $2M$ \gls{p2p} links directly to the co-processors would most likely exceed the number of maximum \gls{p2p} links for any reasonable number of \gls{mami} antennas.
To reduce this number, data can be aggregated using the concept of grouping. 
The different data streams from several \glspl{sdr} are interleaved on one common \gls{sdr} and then sent via one \gls{p2p} link.
Therefore, subsystems are defined, each containing $n_\text{sub}$ \glspl{sdr}.
Data from all antennas within a subsystem is aggregated/disaggregated on the outer two \glspl{sdr} and distributed to the $n_\text{co}$ co-processors using high-speed routers.

At closer look, \figurename~\ref{fig:subsystem} reveals that the \glspl{sdr} on the outer edges which realize the $(n_\text{ant}n_\text{sub}) \ \text{to}\ (n_\text{co})$ and $(n_\text{co}) \ \text{to}\ (n_\text{ant}n_\text{sub})$ router functionalities, require the highest number of \gls{p2p} links, and thus have to deliver the highest throughput.
Hence, the following inequalities have to be fulfilled for the subsystems not to exceed the constraints for maximum number of \gls{p2p} links ($\text{P2P}_\text{SDR,max}$) and maximum bidirectional throughput ($R_{\text{SDR}_\text{max}}$):
\begin{align}
R_{\text{SDR}_\text{max}} > R_{\text{SDR}_\text{out}}=R_{\text{SDR}_\text{in}}=n_\text{ant}\cdot n_\text{sub}\cdot w\cdot F_\text{sub} \label{eq:R_sdr}\\
\text{P2P}_\text{SDR,max} > \text{P2P}_\text{SDR} = n_\text{co}+n_\text{sub}\label{eq:P2P_sdr}
\end{align}
where it is assumed that if an \gls{sdr} employs more than one antenna, the data is interleaved before it is sent to the router on the outer \glspl{sdr}.
The constraints given in equation (\ref{eq:R_sdr})-(\ref{eq:P2P_sdr}) can be used to determine the maximum number of \glspl{sdr} per subsystem ($n_\text{sub}$) such that hardware constraints are not exceeded. 

\subsubsection{Co-processors} 
\begin{figure}
	\centering
	\hspace*{0pt}\input{co_processor.tikz}
	\caption{Shuffling data from the $\lceil M/(n_\text{sub}n_\text{ant})\rceil$ subsystems to the $n_\text{co}$ co-processors. The routers use a simple round robin scheme to combine/distribute the data from/to corresponding subsystems.}
	\label{fig:Co_processors}
\end{figure}
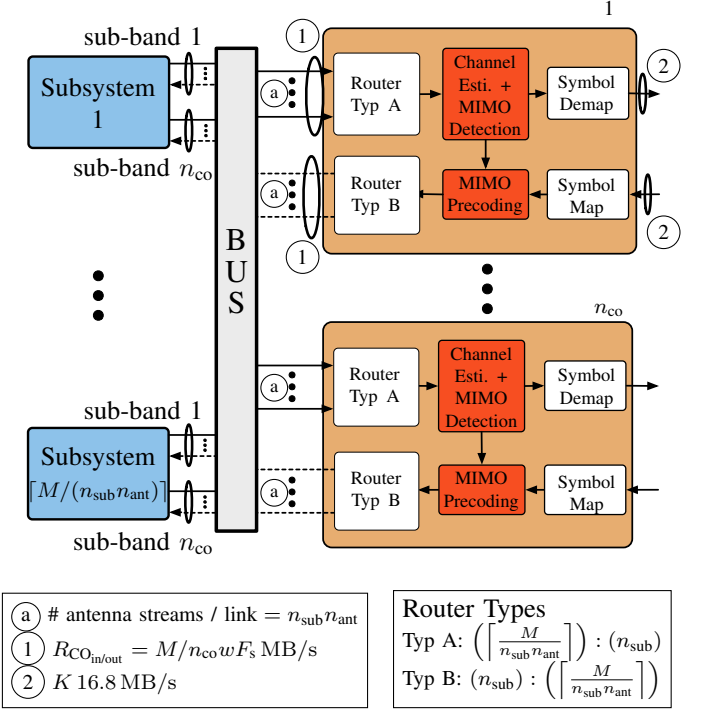

As shown in \figurename~\ref{fig:Co_processors}, detection, precoding, \gls{csi} acquisition, symbol mapping and symbol demapping are integrated in the centrally localized co-processor modules which collect data from all \glspl{sdr}. 
Using \gls{csi} estimated from \gls{ul} pilots, \gls{mimo} processing as discussed in Sec.~\ref{sec:MaMi_Theory} and symbol mapping/de-mapping is performed.  

Based on the selected \gls{ofdm} modulation scheme the subcarrier independence can be exploited allowing each of the $n_\text{co}$ co-processors to work on a sub-band of the overall \SI{20}{\mega\hertz} bandwidth.
This efficiently circumvents issues with throughput and latency constraints in the \gls{mimo} signal processing chain.
The co-processors aggregate/disaggregate data from all the antennas in the system using reconfigurable high-speed routers, as shown in \figurename~\ref{fig:Co_processors} for a system having $\lceil M/(n_\text{sub}n_\text{ant})\rceil$ subsystems and $n_\text{co}$ co-processors.

Similarly to the \glspl{sdr}, the two main constraints for the co-processors are the maximum number of \gls{p2p} links denoted $\text{P2P}_\text{CO,max}$ and the maximum throughput denoted $R_{\text{CO}_\text{max}}$.	

The following inequalities have to hold for the co-processor not to exceed these constraints:
\begin{eqnarray}
R_{\text{CO}_\text{max}} &>& R_{\text{CO}_\text{out}}=R_{\text{CO}_\text{in}}=\nonumber\\
&=&\left(\frac{M\cdot w+K}{n_\text{co}}\right)\cdot F_\text{sub}\label{eq:R_co}\\
\text{P2P}_\text{CO,max} &>& \text{P2P}_\text{CO} = 2\cdot \lceil M/n_\text{sub}\rceil + 2. \label{eq:P2P_co}
\end{eqnarray}

Using this modular and generic system partitioning, \gls{hw} platforms built using modular components can be evaluated.
Note, that expressions (\ref{eq:R_sdr}) - (\ref{eq:P2P_co}) may also be used with other system parameters, \eg by redefining $F_\text{s}$ and $F_\text{sub}$.

\begin{table*}[t]
	\centering
	\footnotesize
	\caption{Selected Hardware from National Instruments}
	\noindent\begin{tabular}{lll}
		\toprule
		\textbf{Type} & \textbf{Model} & \textbf{Features} \\
		\midrule
		\multirow{2}{*}{Host}&\multirow{2}{*}{PXIe-8135} & 2.3 GHz Quad-Core PXI Express Controller \\
		&& Up to 8 GB/s system and 4 GB/s slot bandwidth \\
		\midrule
		\multirow{3}{*}{SDR}&\multirow{3}{*}{USRP RIO 294xR / 295xR} & 2 RF Front Ends and 1 Xilinx Kintex-7 \gls{fpga} \\
		&& Center frequency variable from \SI{1.2}{\giga\hertz} to \SI{6}{\giga\hertz} \\
		&& \SI{830}{\mega B\per\second} bidirectional throughput on up to 15 DMA channels\\
		\midrule
		\multirow{2}{*}{Co-Processor}&\multirow{2}{*}{FlexRIO 7976R} & 1 Xilinx Kintex-7 410T \gls{fpga} \\
		&& \SI{2.4}{\giga B\per\second} bidirectional throughput on up to 32 DMA channels\\
		\midrule
		\multirow{4}{*}{Switch}&\multirow{4}{*}{PXIe-1085} & Industrial form factor 18-slot chassis \\
		&& \SI{7}{\giga B\per\second} bidirectional throughput per slot\\
		&& 2 switches per chassis with inter-switch traffic up to \SI{3.2}{\giga B\per\second} \\
		&& Links between chassis bound to \SI{7}{\giga B\per\second} bidirectional \\
		\midrule
		\multirow{3}{*}{Expansion Module}&\multirow{3}{*}{PXIe-8374} & PXI Express (x4) Chassis Expansion Module \\
		&& Software-transparent link without programming\\
		&& Star, tree, or daisy-chain configuration\\
		\midrule
		\multirow{2}{*}{Reference Clock Source}&\multirow{2}{*}{PXIe-6674T} & \SI{10}{\mega\hertz} reference clock source with $<\SI{5}{ppb}$ clock accuracy \\
		&& 6 configurable I/O connections\\
		\midrule
		Ref. Clock Distribution&OctoClock & \SI{10}{\mega\hertz} 8-channel clock and timing distribution network \\
		\bottomrule
	\end{tabular}
	\label{table:Selected_Hardware}
\end{table*}

\section{\Gls{lumami} Testbed Implementation}
\label{sec:lumami_testbed_implementation}
In this section the \gls{lumami} specific implementation details are discussed based on the aforementioned general architecture.
The \gls{lumami} system was designed with $100$ \gls{bs} antennas and can serve up to $12$ \glspl{ue} simultaneously.
Based on these parameters, the selected modular \gls{hw} platform is presented and given constraints are evaluated.
Consequently, the specific frame structure and other features of the system including base-band processing, antenna array, mechanical structure and synchronization are briefly described.
Before providing details, the authors would like to emphasize, that this is the initial version of the \gls{lumami} testbed and that add-ons and further improvements are planned for the future.

\subsection{Selected Hardware Platform}
The hardware platform was selected based on requirements discussed in Sec.~\ref{sec:System_Design_Aspect}.
Table~\ref{table:Selected_Hardware} shows the selected off-the-shelf modular hardware from National Instruments used to implement the \gls{lumami} testbed.
The \glspl{sdr} \cite{usrp} allow up to 15 \gls{p2p} links ($\text{P2P}_\text{SDR,max}=15$) with a bidirectional throughput of $R_{\text{SDR}_\text{max}}=\SI{830}{\mega B\per\second}$, support a variable center frequency from  \SI{1.2}{\giga\hertz} to \SI{6}{\giga\hertz} and have a \gls{tx} power of \SI{15}{dBm}.
Each \gls{sdr} contains two \gls{rf} chains, \ie $n_\text{ant}=2$, and a Kintex-7 \gls{fpga}.
Selected co-processors \cite{flex} allow a bidirectional \gls{p2p} rate of $R_{\text{CO}_\text{max}}=\SI{2.4}{\giga B\per\second}$ with up to $\text{P2P}_\text{CO,max}=32$ \gls{p2p} links and employ a powerful Kintex-7 \gls{fpga} with a reported performance of up to \SI{2,845}{GMACs\per\second} \cite{Kintex7}.
This is sufficient for a 100 \gls{bs} antenna \gls{mami} testbed due to the fact that $n_\text{co}$ co-processors can be utilized in parallel.
Interconnection among devices is achieved using 18-slot chassis \cite{chassis} combined with per-slot expansion modules \cite{expansion_module}.
Each chassis integrates two switches based on \gls{pcie} using \gls{dma} channels which allow inter-chassis traffic up to \SI{7}{\giga B\per\second} and intra-chassis traffic up to \SI{3.2}{\giga B\per\second}.

The host \cite{host} is an integrated controller, running LabVIEW on a standard Windows operating system and is used to configure and control the system.
The integrated hardware/software stack provided by LabVIEW provides the needed reconfigurability as it abstracts the \gls{p2p} link setup, communication among all devices and allows \gls{fpga} programming as well as host processing using a single programming language. 
An additional feature of LabVIEW is the possibility to seamlessly integrate \gls{ip} blocks generated via Xilinx Vivado platform paving a way to test in-house developed \gls{ip}.  

To be able to synchronize the full \gls{bs}, a Reference Clock Source \cite{timing_mod} and Reference clock distribution network \cite{clock_distribution} are required. 
Their functionalities will be later discussed when presenting the overall synchronization method.

\subsection{Subsystems and Number of Co-processors}
To build the \gls{lumami} testbed with $M=100$ antennas, $50$ \glspl{sdr} are necessary.
The maximum possible subsystem size is chosen to minimize the utilization of available \gls{p2p} links at the co-processors.
By using (\ref{eq:R_sdr}) and an internal fixed-point wordlength of $w=3$ corresponding to a 12-bit resolution on the I- and Q-components, $n_\text{sub}$ is found to be $8$.
As this is not an integer divider of $50$, the last subsystem only contains two \glspl{sdr}.

Based on Table~\ref{tab:data_shuffling_requirements}, the combined subcarrier rate for all antennas is $wMF_\text{sub}=\SI{5}{\giga B\per\second}$ and another $K\cdot F_\text{sub}=\SI{200}{\mega B\per\second}$ are needed for information symbols.
To not exceed $R_{\text{CO}_\text{max}}$ at least three co-processors must be utilized.
To further lower the burden on the design of the low-latency \gls{mimo} signal processing chain, $n_\text{co}=4$ is chosen such that each co-processor processes $300$ of the overall 1200 subcarriers.

Table~\ref{tab:lumami_parameters} summarizes the \gls{lumami} testbed parameters and shows that constraints are met according to (\ref{eq:R_sdr})-(\ref{eq:P2P_co}).
It can also be seen that the design is still within the constraints if scaling up the number of \gls{bs} antennas to $M=128$, which has been done in subsequent designs based on the same hardware, \eg \cite{Bristol1}.

\begin{table}[t]
	\centering
	\footnotesize
	\caption{System Parameters and validation of constraints in the LuMaMi testbed.}
	\noindent\begin{tabular}{llll}
		\toprule
		\multicolumn{2}{c}{\textbf{Parameters}} & \textbf{Rates \SI{}{\mega B\per\second}} \\
		\midrule
		$M$ & $100$ & $R_{\text{SDR}_\text{max}}=830 > R_{\text{SDR}_\text{out}}=R_{\text{SDR}_\text{in}}=806.4$ \\
		$K$ & $12$ & $R_{\text{CO}_\text{max}}=2,400 > R_{\text{CO}_\text{out}}=R_{\text{CO}_\text{in}}=1,460$\\
		\cmidrule(lr){3-3}
		$n_\text{ant}$ & $2$ & \textbf{P2P Links} \\
		\cmidrule(lr){3-3}
		$n_\text{sub}$& $8^\text{a}$ & $\text{P2P}_\text{SDR,max}=15 > \text{P2P}_\text{SDR} = 12$\\
		$n_\text{co}$& $4$ & $\text{P2P}_\text{CO,max}=32 > \text{P2P}_\text{CO} = 18$ \\
		\bottomrule
	\end{tabular}
	\begin{tablenotes}
		\small
		\item[a] $^\text{a}$ Note, that the last subsystem only consists of two SDRs.
	\end{tablenotes}
	\label{tab:lumami_parameters}
\end{table}
\begin{figure}[!t]
	\centering
	\includegraphics[width=0.9\columnwidth]{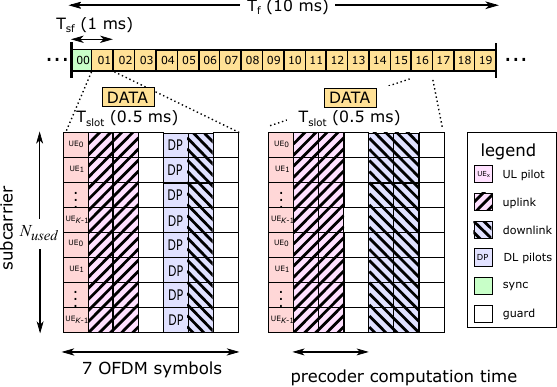}
	\caption{The default frame structure used in the \gls{lumami} testbed.}
	\label{fig:frame_structure}
\end{figure} 
\begin{figure*}
	\center
	\begin{tabular}{cc}
		$\hspace{-.3cm}\vcenter{\hbox{\includegraphics[width=0.8\columnwidth]{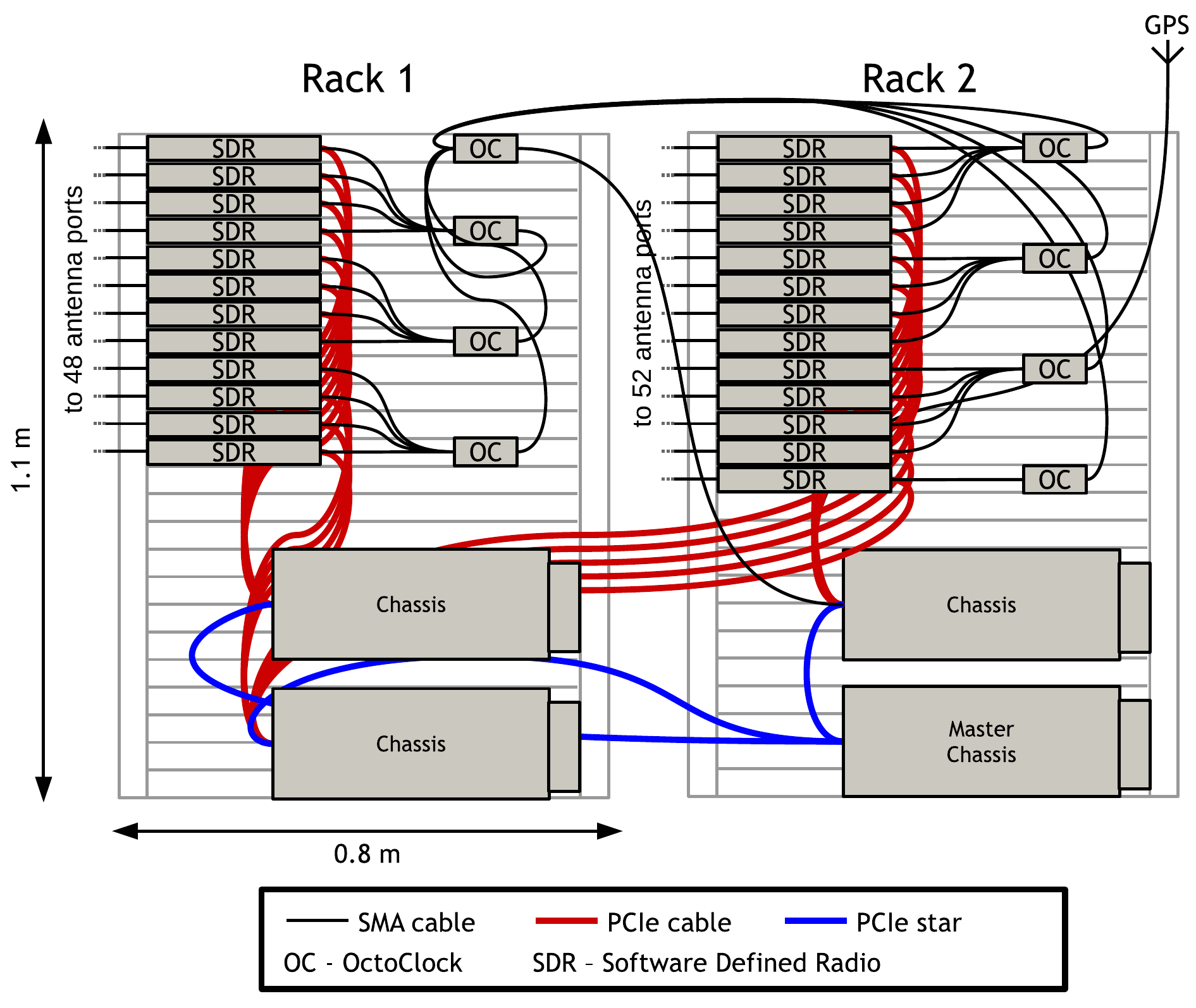}\hspace{1.5cm}}}$
		$\vcenter{\hbox{\includegraphics[width=.5\columnwidth]{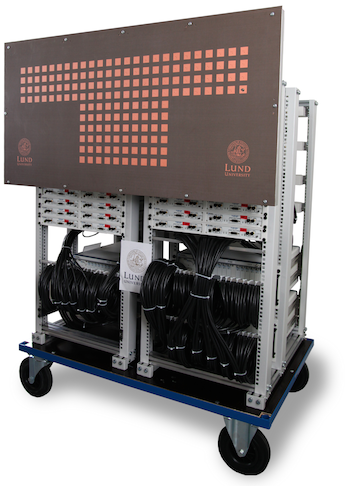}}}$
	\end{tabular}
	\caption{Left: Side view of the mechanical assembly of the \gls{bs}. The two racks sit side by side (not as shown) with the \glspl{sdr} facing the same direction (towards the antenna array). Two columns of USRP \glspl{sdr} are mounted in each rack, totaling 50 of them. Right: The assembled \gls{lumami} testbed at Lund University, Sweden.}
	\label{fig:RackMountedSystem}
\end{figure*}
\subsection{Frame Structure}
The default frame structure for the \gls{lumami} testbed is shown in \figurename~\ref{fig:frame_structure}.
One frame is $T_\text{f}=\SI{10}{\milli\second}$ and is divided in ten subframes of length $T_\text{sf}=\SI{1}{\milli\second}$. 
Each subframe consists of two slots having length $T_\text{slot}=\SI{0.5}{\milli\second}$, where the first subframe is used for control signals, \eg to implement over-the-air synchronization, \gls{ul} power control and other control signaling.
The 18 slots in the other nine subframes encapsulate seven \gls{ofdm} symbols each.
Comparing to \figurename~\ref{fig:frame_structure_generic}, a reciprocity calibration cycle is defined over the whole run-time of the \gls{bs} for simplicity and due to the fact that there is no large drift after warming up the system in a controlled environment\cite{Vieira2014a}.
The \gls{dl} pilot cycles and control cycles are both set to be the length of one frame.
Each frame starts with one control subframe followed by one subframe with one \gls{dl} pilot and one \gls{dl} data symbol whereas all others use two DL data symbols.

\subsection{Mobility}
The pilot distance in time in the default frame structure given in \figurename~\ref{fig:frame_structure_generic} is $T_\text{p}\approx \SI{430}{\micro\second}$ or six \gls{ofdm} symbols.
Thus, $\nu_\text{max}\approx\SI{240}{\hertz}$ for a correlation of 0.9.
Due to availability from a network operator, a carrier frequency of $f_\text{c}=\SI{3.7}{\giga\hertz}$ is selected. 
Using (\ref{eq:Mobility}), $\text{v}_\text{max}=\SI{70}{\kilo\meter\per\hour}$ is found as maximum supported speed.

\subsection{\gls{tdd} Turnaround Time}
The pre-coding turnaround time requirement for the implementation can be analyzed based on (\ref{eq:latency}).
The analog front-end delay of the \glspl{sdr} was measured to be about $\SI{2.25}{\micro\second}$.
Taking the frame structure in \figurename~\ref{fig:frame_structure} (assuming $\Delta^\text{rf,\tiny{TX}}=\Delta^\text{rf,\tiny{RX}}$ which is not necessarily true), the latency budget for base-band processing is as follows:
Overall time for pre-coding after receiving the UL pilots is $\SI{214}{\micro\second}$ ($3$ \gls{ofdm} symbols).
The 2048 point \gls{fft}/\gls{ifft} (assuming a clock frequency of \SI{200}{\mega\hertz}) requires around $\SI{35}{\micro\second}\times 2 = \SI{70}{\micro\second}$ in total for TX and RX (including sample reordering). 
As a result, the remaining time for channel estimation, \gls{mimo} processing, and data routing is around $\SI{140}{\micro\second}$, which is the design constraint for this specific frame structure.

An analysis of the implemented design showed that the latency is far below the requirement for the default frame structure which makes it possible to use the testbed for higher mobility scenarios from this point of view\cite{sips_steffen}.

\subsection{Implementation Features}
\subsubsection{Base-band Processing}
On the \gls{lumami} testbed, each \gls{ue} sends pilots on orthogonal subcarriers, \ie each \gls{ue} uses every $K$-th subcarrier with the first \gls{ue} starting at subcarrier 0, the second at subcarrier 1 etc., overall utilizing a full \gls{ofdm} symbol.
It was shown that performance does not suffer significantly compared to a full detector calculated for each subcarrier using this method\cite{deliverable3_2}.
Moreover, it efficiently remedies processing requirements and reduces the required memory for storing estimated \gls{csi} matrices by a factor of $K$.
A least-square \gls{csi} estimation algorithm with zeroth-order hold over $K=12$ subcarriers was implemented, however, better estimates could be obtained by on-the-fly interpolation between the estimated subcarriers.
Overall, utilizing this approach reduces the required detection matrix throughput to one matrix every 12 subcarriers, \ie \SI{16.8e6}{subcarriers\per\second/12} = \SI{1.4e6}{Detection Matrices\per\second}.  

Two versions for detection were implemented.
The first one based on a QR decomposition of the channel matrix augmented with the regularizations factors to a matrix of size $2M\times K$.
This is then formulated into a partial parallel implementation employing a systolic array\cite{1285069}.
The latter one based on a Neumann-series\cite{Prabhu2013}.
In the QR decomposition, each column is processed using the discrete steps of the modified Gram-Schmidt algorithm.
The logic on the co-processors can be reconfigured so that the same hardware resources that provide the \gls{rzf} decoder can also provide the \gls{zf} and \gls{mrc} decoders, \ie the detection / precoding schemes discussed in Sec.~\ref{sec:MaMi_Theory} are supported with run-time switching.
The Neumann-series based \gls{zf} detector utilizes the unique property that in \gls{mami}, the Gramian matrix shows dominant diagonal elements if \glspl{ue} use \gls{ul} power control, or if scheduling is performed to serve \glspl{ue} with similar power levels in the same time/frequency block to mitigate the influence of path loss differences. This, allows the matrix inversion to be approximated with low overall error \cite{Prabhu2013}.
The utilizations for the two \gls{fpga} designs are shown in Table~\ref{tab:FPGA_resources}.
\begin{table}[t]
	\centering
	\small
	\caption{FPGA Utilization for two different MIMO processing implementations}
	\noindent\begin{tabular}{ccccc}
		\toprule
		Implementation & Registers & LUT & RAMs & DSP48\\
		\midrule
		\multirow{2}{*}{QRD} & 46470  & 49315  & 171 & 596 \\
		& (9.1\%) & (20.3\%) & (21.5\%) & (38.7\%) \\
		\midrule
		\multirow{2}{*}{Neumann-Series} & 16000  & 28700 & 6 & 176 \\
		& (3.1\%) & (11.8\%) & (0.75\%) & (11.4\%) \\
		\bottomrule
	\end{tabular}
	\label{tab:FPGA_resources}
\end{table}
Clearly, overall processing complexity and resource utilization can be significantly reduced by exploiting the special properties of \gls{mami}.

At this point, the regularizations factors $\beta_{\mathrm{reg}_\mathrm{pre}}$ and $\beta_{\mathrm{reg}_\mathrm{dec}}$ are not run-time optimized but set manually, however, implementation of this feature is planned in future.
For a more detailed discussion of the low-latency signal processing implementation on the testbed we refer to\cite{sips_steffen}.

\subsubsection{Host-based visualization and data capturing} 
The available margin of \SI{1}{\giga B\per\second} and 14 \gls{p2p} links to the corresponding maximum values on the co-processors are used for visualization and system performance metrics.
The host receives decimated equalized constellations and raw subcarriers for one \gls{ul} pilot and one \gls{ul} data symbol per frame.
These features add another 
\begin{equation}
\frac{300\cdot 2\text{bytes}+2\cdot 300\cdot 4\text{bytes}}{\num{10}\text{ms}}=\SI{300}{\mega B\per\second}\nonumber
\end{equation}
of data flowing in and out of the co-processor.
The raw subcarriers are used to perform channel estimation and \gls{ul} data detection on the host computer with floating point precision and allow fast implementation of different metrics, like constellation, channel impulse response, power level per antenna and user.
Another 12 \gls{p2p} links available are utilized to transmit and store real-time \glspl{ber} for all 12 \glspl{ue}.

Moreover, to be able to capture dynamics in the channel for mobile \glspl{ue}, \gls{csi} can be stored on a \SI{}{\milli\second} basis.
An integrated \SI{2}{\giga B} \gls{dram} buffer on each of the co-processors was utilized for this since direct streaming to disk would exceed the \gls{p2p} bandwidth limits.
Snapshots can either be taken for \SI{60}{\second} in a \SI{5}{\milli\second} interval or over \SI{12}{\second} in a \SI{1}{\milli\second} interval, both corresponding to \SI{2}{\giga B} of data for $300$ subcarriers per co-processor.
\subsubsection{Scalability/Reconfigurability}
Before startup, the number of deployed \gls{bs} antennas can be arbitrarily set between $4$ and $100$.
This is achieved by introducing zeros for non-existing antennas within the \gls{lut}-based reconfigurable high-speed routers on the co-processors, thereby allowing to evaluate effects of scaling the \gls{bs} antennas in real environments\cite{sips_steffen}.
Additionally, all $140$ \gls{ofdm} symbols in a frame can be rearranged arbitrarily before start-up while each frame always repeats itself.
For instance, we can choose to set the first symbol as UL pilots and all others as UL data in a static UL only scenario.

\subsubsection{Reciprocity Calibration}
Estimation of the reciprocity calibration coefficients was implemented on the host, mainly for two reasons: \emph{(i)} the host can perform all operations in floating-point which increases precision and \emph{(ii)} the drift of the hardware is not significant once the system reached operating temperature \cite{Vieira2014a}.
Estimated reciprocity coefficients are applied in a distributed manner on the \glspl{sdr}\cite{sips_steffen}.

\subsection{Mechanical structure and electrical characteristics}
\label{subsec:MecDesign}
Two computer racks containing all components measuring $0.8 \times 1.2 \times 1$\,m were used, as shown \figurename~\ref{fig:RackMountedSystem}.
An essential requirement for the \gls{lumami} testbed is to allow tests in different scenarios, \eg indoor and outdoor.
Therefore, the rack mount is attached on top of a 4-wheel trolley.

\subsection{Antenna Array}
\label{subsec:AntArray}
The planar T-shaped antenna array with 160 dual polarized $\lambda$/2 patch elements was developed in-house.
A 3.2\;mm Diclad 880 was chosen for the printed circuit board substrate. 
The T upper horizontal rectangle has $4\times25$ elements and the central square has $10\times10$ elements (see \figurename~\ref{fig:RackMountedSystem} right).
This yields 320 possible antenna ports that can be used to explore different antenna array arrangements, for example $10\times 10$ or $4\times 25$ with the latter one being the default configuration.
All antenna elements are center shorted, which improves isolation and bandwidth.
The manufactured array yielded an average \SI{10}{\decibel}-bandwidth of \SI{183}{\mega\hertz} centered at \SI{3.7}{\giga\hertz} with
isolation between antenna ports varying between \SI{18}{\decibel} and \SI{28}{\decibel} depending on location in the array.

\subsection{User Equipment}
Each \gls{ue} represents a phone or other wireless device with single antenna capabilities.
One \gls{sdr} serves as two independent \glspl{ue} such that overall six \glspl{sdr} are required for the 12 \glspl{ue}.
The base-band processing, \ie \gls{ofdm} modulation/demodulation and symbol mapping/demapping are essentially identical to the \gls{bs} implementation.
A least-square \gls{csi} acquisition is performed on precoded DL pilot followed by a \gls{zf}-equalizer.
The \gls{dl} pilots occupy a full \gls{ofdm} symbol.
The \glspl{ue} may be equipped with any type of antenna using SMA connectors.
\subsection{Synchronization}
A \gls{mami} \gls{bs} requires time synchronization and phase coherence between each RF chain.
This is achieved using the \SI{10}{\mega\hertz} reference clock source and the reference clock and trigger distribution network (see Table~\ref{table:Selected_Hardware}).
The reference clock is used as the source of each radio local oscillator, providing phase coherence among devices.
The trigger signal is used to provide a time reference to all the radios in the system. 
A master provides an output digital trigger that is amplified and divided among all the radios.
Upon receipt of the rising edge of the event trigger, all \glspl{sdr} are started.
The basic structure can be identified in \figurename~\ref{fig:RackMountedSystem} on the left.

To synchronize the \glspl{ue} with the \gls{bs} \gls{ota}, the \gls{lte} Zadoff-Chu \gls{pss} is used, which occupies the center \SI{1.2}{\mega\hertz} of the overall bandwidth.
\gls{ota} synchronization and frequency offset compensation are achieved by employing a frequency-shifted bank of replica filters.
The process follows a two step procedure: finding a coarse candidate position by scanning over the whole radio frame followed by tracking the \gls{pss} in a narrowed window located around the coarse candidate position.
Additionally, by disciplining the \gls{ue} \glspl{sdr} with \gls{gps}, frequency offset compensation may be avoided by lowering the frequency offset to $<\SI{300}{\hertz}$.

\begin{figure}[!t]
	\centering
	\includegraphics[width=.85\columnwidth]{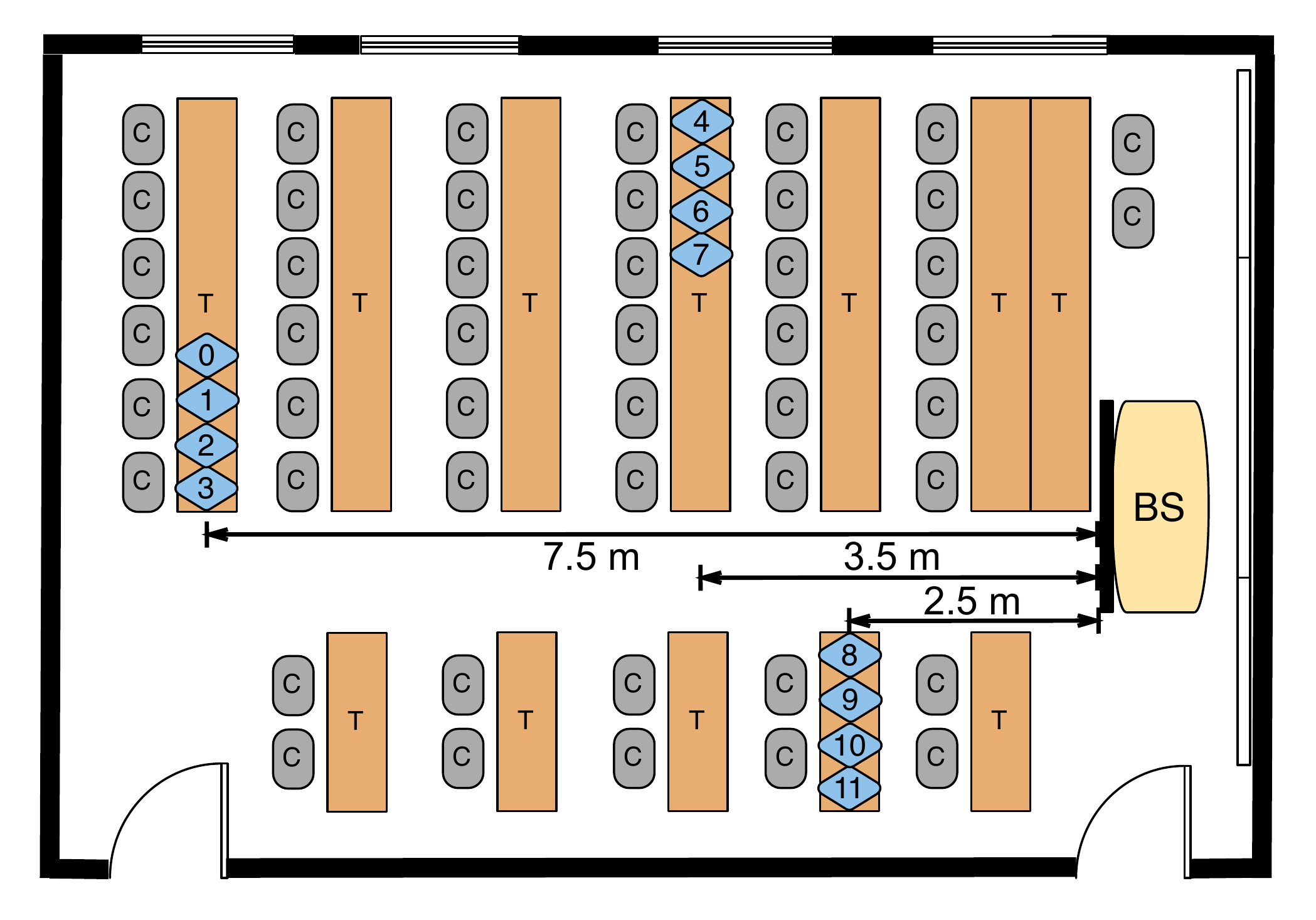}
	\caption{The indoor measurement setup in a lecture room including the positions of the 12 \glspl{ue}. The \gls{bs} is shown at the right-hand side and is situated at the front of the lecture hall. The terminals are placed in groups of four on three different tables and distances to the \gls{bs}.}
	\label{fig:room2311}
\end{figure} 
\begin{figure}[!t]
	\centering
	\includegraphics[width=0.85\columnwidth]{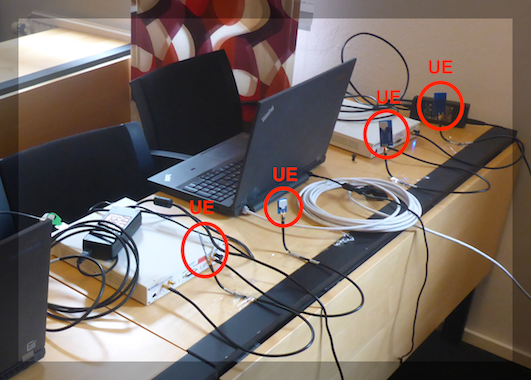}
	\caption{One group of four \glspl{ue} with a high user density per unit area to validate the spatial multiplexing capabilities of \gls{mami}.}
	\label{fig:indoor_UE_group}
\end{figure} 
\begin{figure*}[!t]
	\centering
	\input  {Grouped_BERs.tikz}%
	\caption{UL and DL \glspl{ber} for 12 \glspl{ue} with \gls{zf} decoder/precoder.}
	\label{fig:BERs}
\end{figure*}
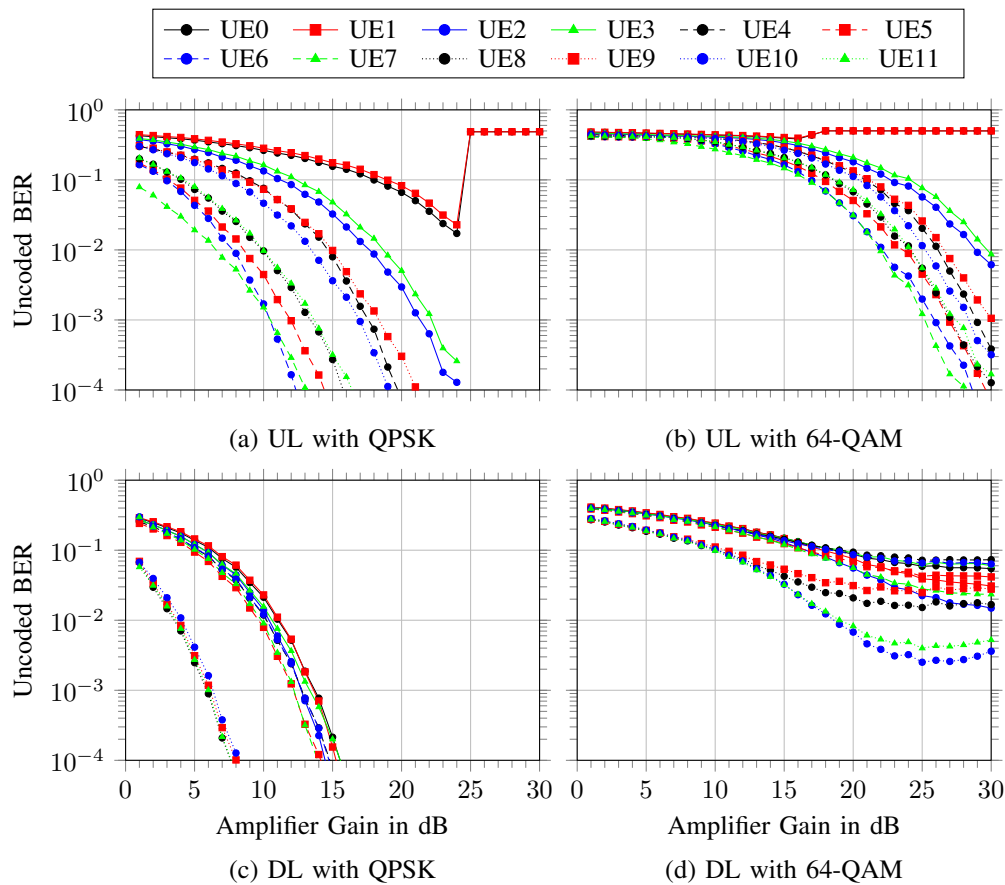 

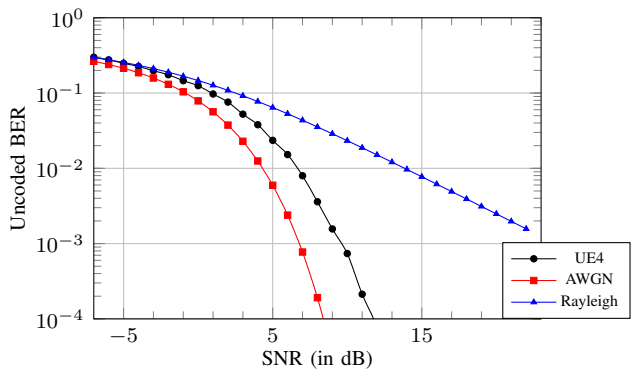
\begin{figure}[!t]
	\centering
	\def\filename{UL_all_users_qpsk_with_awgn_user_4.dat}
	\input  {BER_plot_AWGN.tikz}%
	\caption{Comparing the \gls{ber} of \gls{ue}4 to \gls{awgn} and Rayleigh fading channels.}
	\label{fig:BER_AWGN}
\end{figure} 

\section{Proof-of-concept Results}
\label{sec:Proof_of_Concept_and_Results}
This section describes two experiments performed to validate our testbed design, the \gls{mami} concept and its performance. 
The first test is performed indoors with high density of users per area unit to stress the spatial multiplexing capabilities of the system. 
The second test is conducted outdoors with less dense deployment of \glspl{ue} and is primarily designed to test the range and multiplexing capabilities outdoors.
For all tests, the default antenna configuration, \ie $4\times 25$ was used on the \gls{bs} side whereas the \glspl{ue} were equipped with linear polarized ultra-wideband antennas.
It has to be noted that all results shown in this section are obtained from real-time operation without \gls{ul} power control.

\subsection{Indoor Test}
In this test real-time uncoded BER curves are measured, employing \gls{mrc}/\gls{mrt} and \gls{zf} as decoders/precoders.
The \gls{ul} \gls{ber} curves are obtained by sweeping all \gls{ue} \gls{tx} \gls{pa} gains synchronously, and for the \gls{dl} \gls{ber} curves the \gls{pa} gains of the \gls{bs} \gls{tx} chains while keeping other system parameters constant.
Note that the initial parameterization of the system is chosen empirically, so it allows smooth \gls{ber} curves starting at about $0.5$. 
Each gain step is held constant for about \SI{4}{\second} corresponding to about \num{36e6} and \num{108e6} transmitted bits per step for QPSK and 64-QAM modulation, respectively.

\subsubsection{Scenario}
Twelve \glspl{ue} are set up in a lecture hall at Lund University with the \gls{bs} at the front as shown in \figurename~\ref{fig:room2311} including the respective \gls{ue} placements.
All \glspl{ue} are packed in groups of four resulting in a high density of \glspl{ue} per area unit.
One of these groups can be seen in \figurename~\ref{fig:indoor_UE_group}.
\subsubsection{UL \glspl{ber}}
\label{subsubsec:indoor_UL_BERs}
\figurename~\ref{fig:BERs}, (a) and (b), show the \glspl{ber} for all 12 \glspl{ue} using \gls{zf} detector for QPSK and 64-QAM modulation, respectively.
For both constellation sizes, the \glspl{ue} furthest away, \gls{ue}0 to \gls{ue}3 show highest \gls{ber}.
\gls{ue}0 and \gls{ue}1 even show a sudden increase for the \gls{ber} to 0.5 which was diagnosed to be due to saturation of their respective \glspl{pa}.
Moreover, their performance shows severe limitation compared to the other \glspl{ue}, giving a clear indication that their performance is interference rather than power limited.  
The group closest to the \gls{bs}, \gls{ue}9-\gls{ue}12, shows best performance although the variation within the group is still quite significant.
Overall, the expected trend, increasing performance with increased transmit gain is clearly noticeable with the \gls{ber} curve shapes resembling those of \gls{awgn} channels.
Comparing the amplifier gain settings for QPSK and 64-QAM to achieve the same \gls{ber} the differences are found to be in the range of \SI{10}{\decibel} to \SI{16}{\decibel} whereas a difference of \SI{9}{\decibel} is expected for \gls{awgn}.
Overall, it can be seen that all \glspl{ue} except \gls{ue}0 and \gls{ue}1 achieve \gls{ber} below \SI{10}{\percent} at an amplifier gain of \SI{15}{\decibel} for QPSK and \SI{25}{\decibel} for 64-QAM, respectively.

\subsubsection{DL \glspl{ber}}
\figurename~\ref{fig:BERs}, (c) and (d), show the DL \glspl{ber} using \gls{zf} precoder for QPSK- and 64-QAM modulation, respectively.
Using QPSK modulation, the group closest to the \gls{bs}, \gls{ue}9-\gls{ue}12, achieves a considerably better performance than the other two groups.
Using 64-QAM, all \glspl{ue} show an error-floor towards higher \gls{tx} gain values
which is likely a result of imperfect reciprocity calibration combined with leakage among \glspl{ue} due to non-perfect channel knowledge resulting in interference among \glspl{ue}.
However, for the QPSK modulation case all \glspl{ue} experience better \gls{ber} rates which can be explained by the significantly higher available transmit power on the \gls{bs} side, utilizing 100 active \gls{rf}-chains.
Comparing again the difference in amplifier gain setting for QPSK and 64-QAM, their differences are about \SI{12}{\decibel} to \SI{16}{\decibel}.
The tests performed were mainly to prove functionality, and thus, no special care was taken to achieve best possible accuracy for the reciprocity calibration.
However, individual parts are continuously tested to be improved.

\subsubsection{Performance Evaluation}
While the \glspl{ber} plots in Fig.~\ref{fig:BERs} nicely show the trend with increasing transmit power, they do not provide a real performance indication against \gls{snr}. 
The current implementation of the testbed does not provide \gls{snr} estimates in real-time such that the data presented in Fig.~\ref{fig:BERs} can be seen as the raw data provided during measurements.
To provide an indication of the system performance the \gls{snr} of \gls{ue}4 was estimated based on the received \gls{ul} channel estimates.
Estimated subcarriers at different time instances (about \SI{200}{\milli\second} apart) were subtracted / added to extract the noise / signal plus noise level which was then used to calculate the \gls{snr} value.
However, this practice has limits as for close users interference may be stronger than the noise whereas for far away users the signal level may be too low.
Therefore, \gls{ue}4 was chosen which due to its placement during the measurement allowed a relatively good \gls{snr} estimation.
Fig.~\ref{fig:BER_AWGN} shows the \gls{ber} of \gls{ue}4 in comparison with the theoretical performance in \gls{awgn} and Rayleigh fading channels.
It is visible that due to the excess amount of \gls{bs} antennas the performance is close to the \gls{awgn} channel. 
To be more specific, due to the channel hardening the performance is only about \SI{3}{\decibel} worse than for a \gls{awgn} channel which would be achieved for perfect channel hardening.
On the \gls{dl} the \glspl{snr} are affected by several factors including the higher overall transmit power from the 100 active \gls{rf}-chains and possible inaccuracies in the reciprocity calibration coefficients.
As \gls{dl} precoding is performed based on \gls{ul} channel estimates, \gls{snr} estimation is practically not feasible.

As all shown \glspl{ber} curves closely resemble the shape of an \gls{awgn} channel it can be claimed that the \gls{mami} concept works and is capable of serving 12 \glspl{ue} on the same time/frequency resource even with a high \gls{ue} density which in turn significantly improves the spectral efficiency compared to current cellular standards.

\subsubsection{MRC/MRT versus ZF}
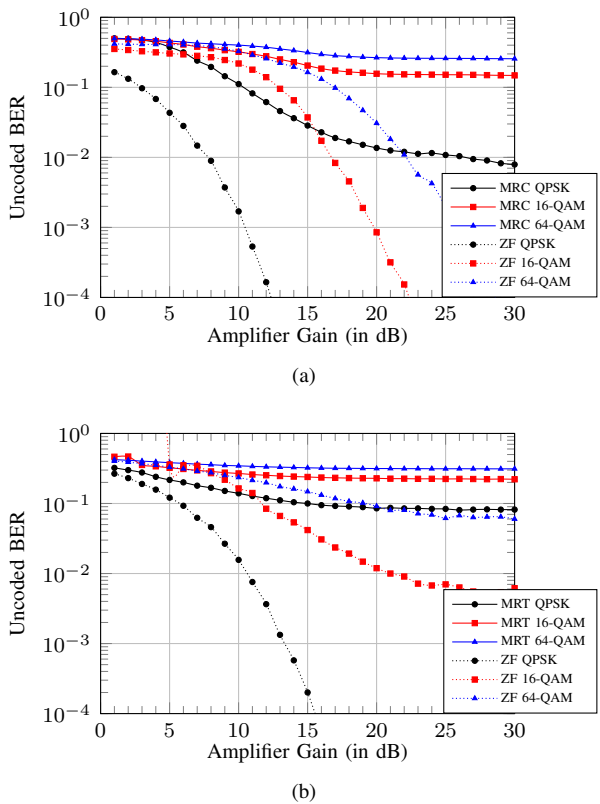
\begin{figure}[!t]
	\centering
	\subfloat[]{\hspace{-0.5cm}
		\def\filename{UL_one_user_three_const_UE11.dat}
		\def\detused{MRC}
		\input  {MRC_vs_ZF.tikz}%
		\label{subfig:UL_both_det}}\\
	\subfloat[]{\hspace{-0.5cm}
		\def\filename{DL_one_user_three_const_UE11.dat}
		\def\detused{MRT}
		\input  {MRC_vs_ZF.tikz}%
		\label{subfig:DL_both_pre}}
	\caption{\glspl{ber} for \glspl{ue}7 using QPSK, 16-QAM and 64-QAM modulation. \protect\subref{subfig:UL_both_det} on the \gls{ul} for \gls{zf} and \gls{mrc} detector and \protect\subref{subfig:DL_both_pre} on the \gls{dl} for \gls{zf} and \gls{mrt} precoder.}
	\label{fig:MRC_vs_ZF}
\end{figure} 
To compare the performance of \gls{mrc}/\gls{mrt} and \gls{zf} it is beneficial to isolate the analysis to one \gls{ue}.
\figurename~\ref{fig:MRC_vs_ZF}a and  \figurename~\ref{fig:MRC_vs_ZF}b show the \gls{ber} for \gls{ue}7 for QPSK, 16-QAM and 64-QAM modulations while the \gls{bs} employs either \gls{mrc}/\gls{mrt} or \gls{zf} on the \gls{ul} and \gls{dl}, respectively.

Overall, \gls{zf} shows an superior performance trend with increasing \gls{pa} gains, while the performance of \gls{mrc} appears to level off\footnote{This is expected from theory, as inter-user interference is the main source of error during data detection. The high density users setup adopted in this experiment highly contributes to this phenomena.}.
Looking in more detail, \gls{zf} is capable of achieving more than an order of magnitude lower \glspl{ber}, compared to \gls{mrc}.
Using higher constellation sizes, 16-QAM or 64-QAM, the results for \gls{mrc} show an even more significant deterioration. 
On the \gls{dl}, \gls{zf} also outperforms \gls{mrt} by far, the latter shows a significant error floor towards higher gains as in the \gls{ul} case.

Unfortunately, direct comparison between \gls{ul} and \gls{dl} results shown here is not easy to perform.
This is due to the fact that on the \gls{ul}, the performance is isolated to the \gls{ul} transmit power only whereas on the \gls{dl} a combination of \gls{ul} channel estimate quality, \gls{dl} transmit power and reciprocity accuracy determines overall performance.

\subsection{Outdoor Test}
\begin{figure}[!t]
	\centering
	\includegraphics[angle=0,width=0.95\columnwidth]{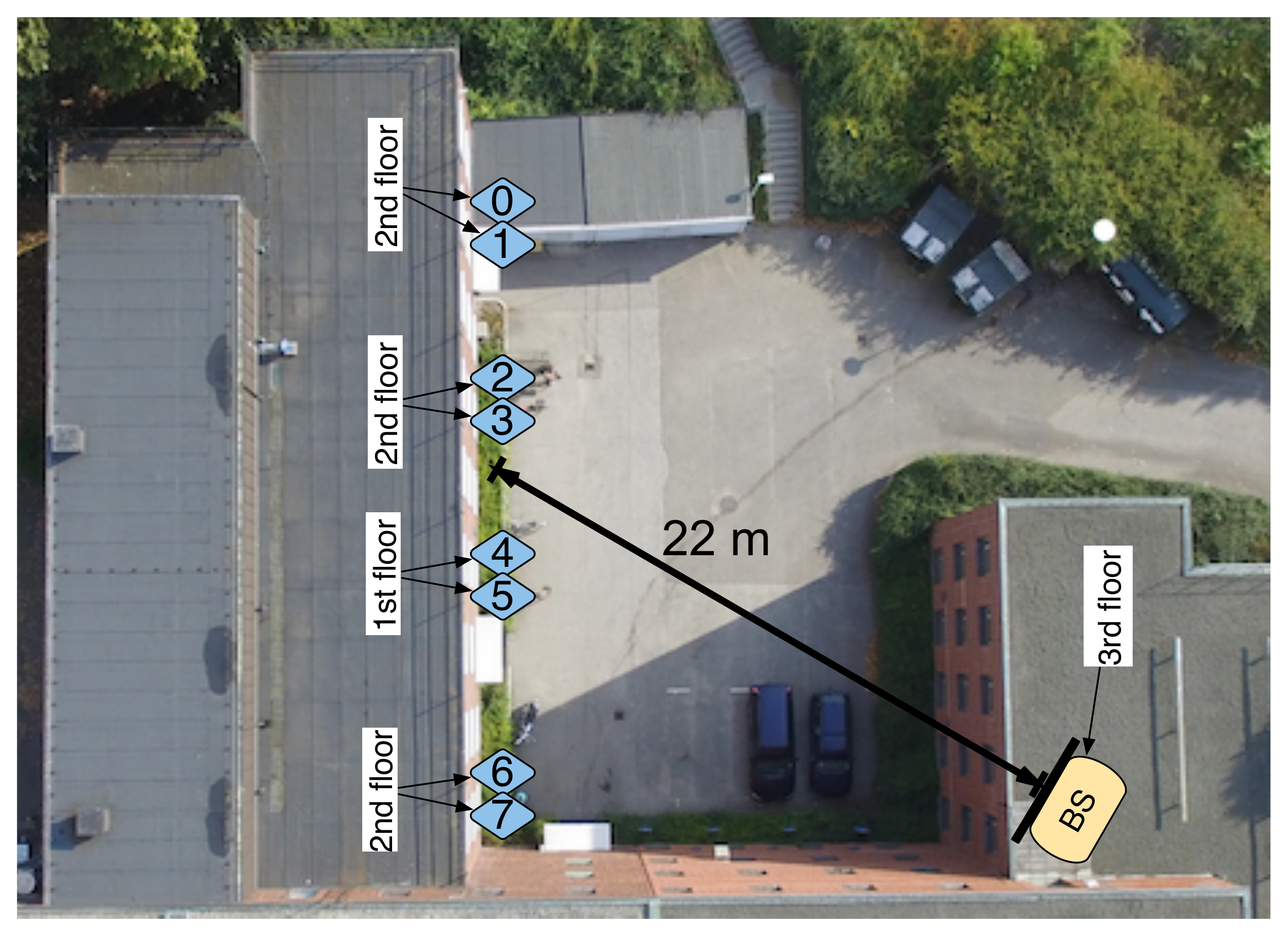}
	\caption{Scenario for the outdoor tests. \gls{bs} placed on the rooftop of the building (third floor) serving eight \glspl{ue} on the opposite wing, with six \glspl{ue} on second floor and two \glspl{ue} on first floor.}
	\label{fig:outdoor_scenario}
\end{figure}

For the outdoor test, the testbed was placed on the rooftop of one of the wings of the department building while the \glspl{ue} where placed on the opposite wing utilizing scaffolding mounted to the building.
Up to eight \glspl{ue} were served simultaneously in a distance of about 18 to 22 meters, six on the second floor and two on the first floor while the testbed was situated on the third floor (rooftop).
The scenario is shown in \figurename~\ref{fig:outdoor_scenario}.

\begin{figure}[!t]
	\centering
	\includegraphics[angle=0,width=0.75\columnwidth]{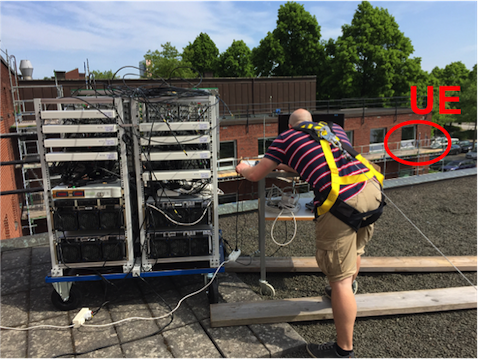}
	\caption{The outdoor test scenario setup with the \gls{bs} deployed on the rooftop of the department building marked with two \glspl{ue} on the opposite building wing.}
	\label{fig:Outdoor_figures}
\end{figure}

\figurename~\ref{fig:Outdoor_figures} shows the \gls{bs} placed on the rooftop of the department building facing towards the opposite wing. The placement for \glspl{ue} 0 and 1 is also marked.

\begin{figure}[!t]
	\centering
	\subfloat[]{\includegraphics[width=.85\columnwidth]{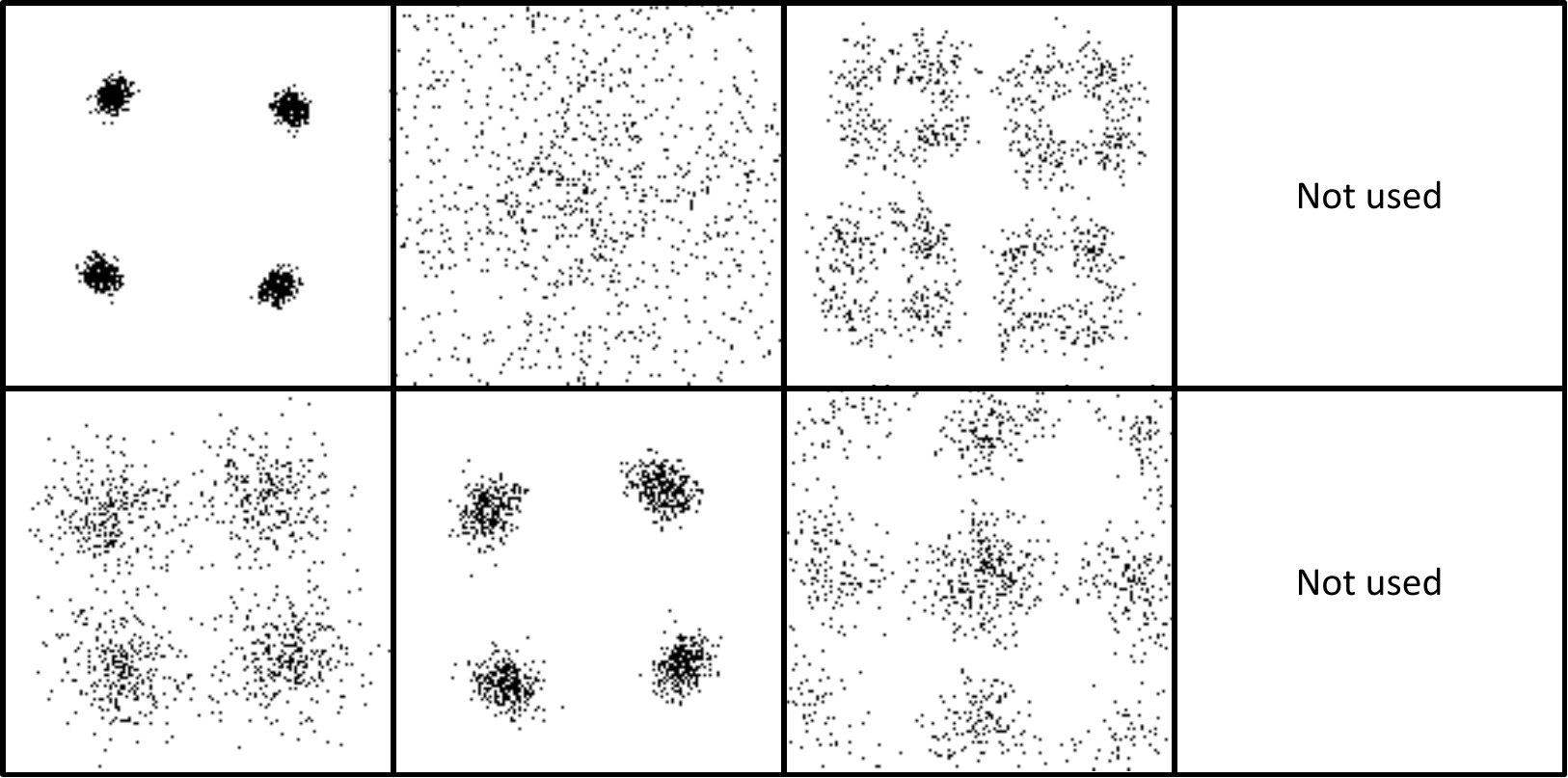}%
		\label{subfig:UL_const_MRC}}\\
	\subfloat[]{\includegraphics[width=.85\columnwidth]{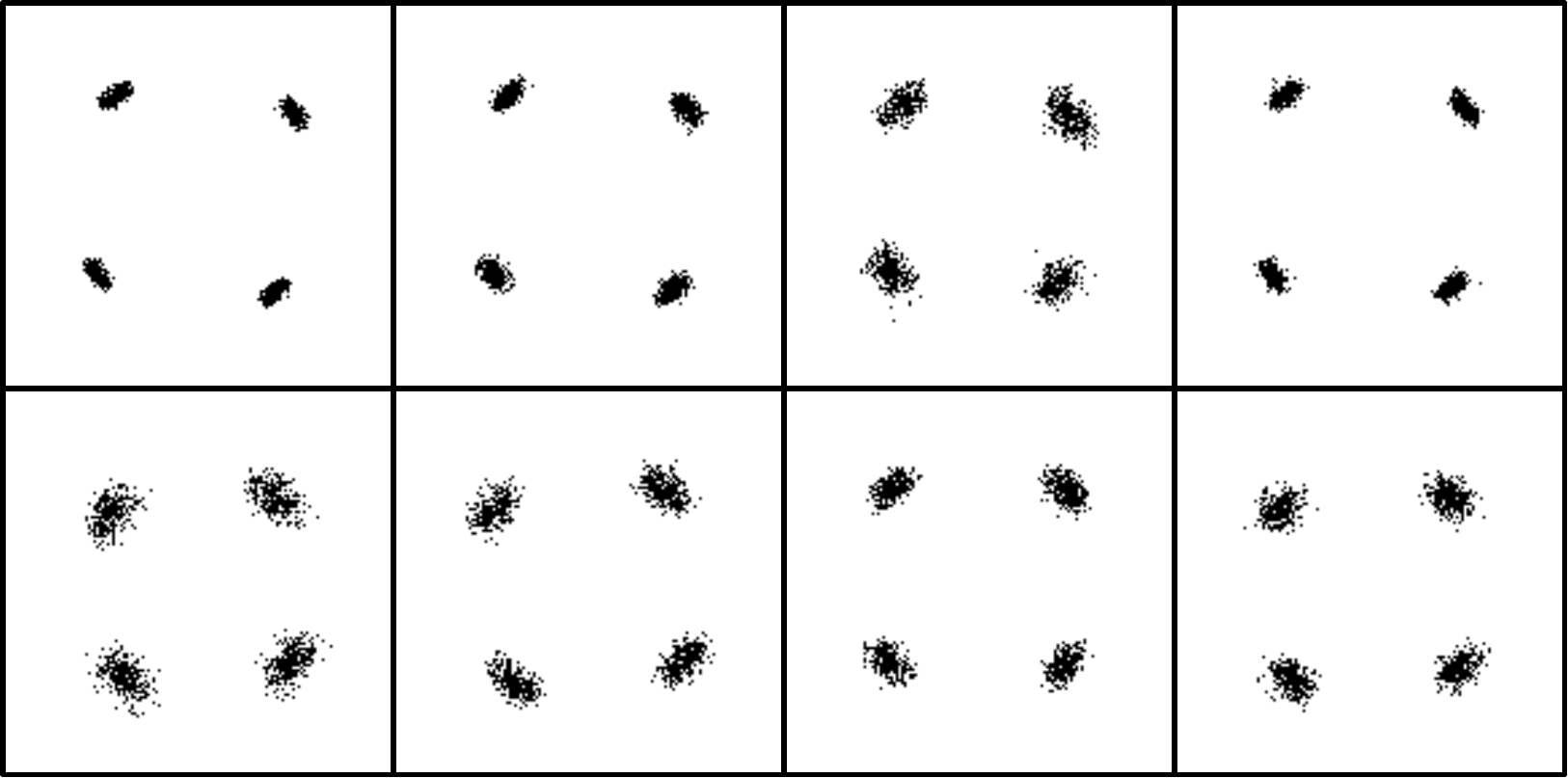}%
		\label{subfig:UL_const_ZF}}
	\caption{\gls{ul} constellations for the outdoor experiment: \protect\subref{subfig:UL_const_MRC} when using \gls{mrc} with 6 \glspl{ue} and \protect\subref{subfig:UL_const_ZF} when using \gls{zf} to serve 8 \glspl{ue}.}
	\label{fig:UL_const_MRC_ZF}
\end{figure}

\figurename~\ref{fig:UL_const_MRC_ZF} shows a screenshot of the received \gls{ul} QPSK constellations for this test setup when using \gls{mrc} and \gls{zf}, respectively.
Using \gls{mrc} without \gls{ecc} for this test, the six \glspl{ue} show significant interference.
Therefore, focus is put on the results obtained with \gls{zf} which is capable of separating up to eight \glspl{ue} and shows very clear constellations, due to the interference suppression.

\begin{figure}[!t]
	\centering
	\savebox{\myimage}{\hbox{\includegraphics[width=0.45\columnwidth]{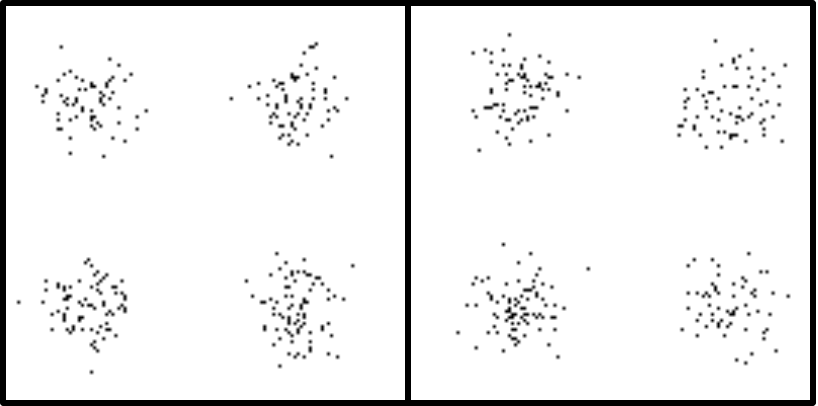}}}%
	\subfloat[]{\usebox{\myimage}
		\label{subfig:T01}}
	\subfloat[]{\raisebox{\dimexpr.5\ht\myimage-.5\height\relax}{\includegraphics[width=0.45\columnwidth]{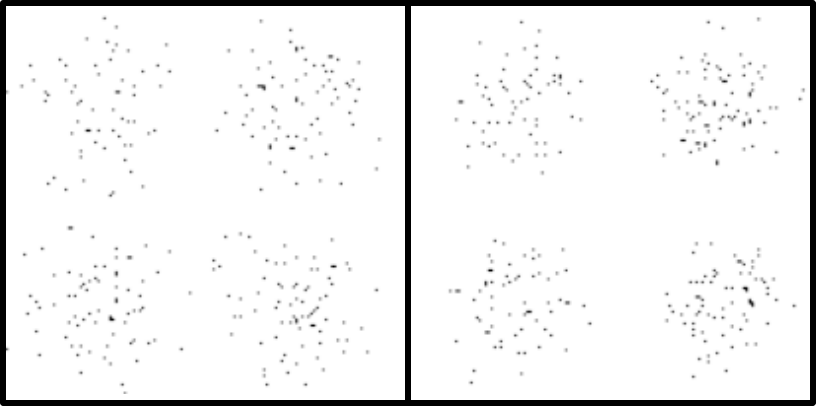}%
			\label{subfig:T02}}}\\
	\hspace{0.1cm}\savebox{\myimage}{\hbox{\includegraphics[width=0.45\columnwidth]{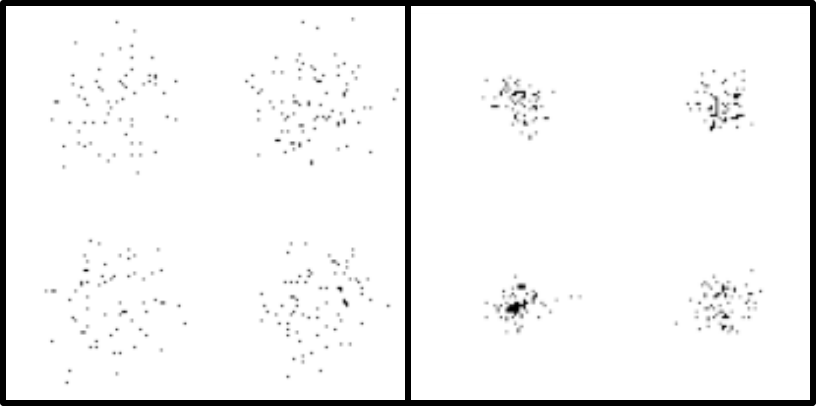}}}%
	\subfloat[]{\usebox{\myimage}
		\label{subfig:T03}}
	\subfloat[]{\raisebox{\dimexpr.5\ht\myimage-.5\height\relax}{\includegraphics[width=0.45\columnwidth]{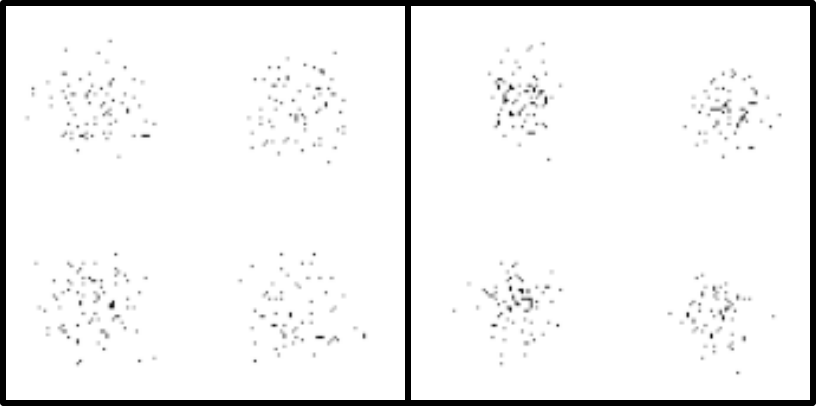}%
			\label{subfig:T04}}}
	\caption{Received DL constellations using \gls{zf}: \protect\subref{subfig:T01} UE0 \& UE1 \protect\subref{subfig:T02} UE2 \& UE3 \protect\subref{subfig:T03} UE5 \& UE8 and \protect\subref{subfig:T04} UE9 \& UE10.}
	\label{fig:DL_const_outdoor}
\end{figure}

Considering \gls{zf} on the \gls{dl}, the constellations for all 8 \glspl{ue} can be seen in \figurename~\ref{fig:DL_const_outdoor}.
Although in-detail analysis is not provided for this test, it is clearly visible that \gls{zf} outperforms \gls{mrc} which is often claimed to be sufficient in literature when analyzing performance based on \gls{iid} channel models\cite{Marzetta2010}.
The results observed in this experiment are representative for most tests performed so far, \ie \gls{dl} always showed to be the more challenging duplex case. 

The \gls{lumami} testbed was also utilized to perform the first \gls{mami} outdoor mobility measurements involving moving pedestrians and cars as \glspl{ue}, however, a discussion of this is out of scope of this paper.
Results and analysis from the mobility tests can be found in\cite{2017arXiv170108818H}.

\section{Conclusion}
\label{sec:conclusion}
This paper presented the \gls{lumami} testbed, which is the first fully operational real-time testbed for prototyping massive MIMO.
Based on massive MIMO system requirements, system parameters were discussed and defined.
Further, a detailed generic hardware partitioning to overcome challenges for data shuffling and \glsentrylong{p2p} link limitations while still allowing scalability, was proposed.
By grouping \glsentrylongpl{sdr} and splitting overall bandwidth, implementation of massive MIMO signal processing was simplified to cope with challenges like \glsentrylong{tdd} precoding turnaround time and limited \glsentrylong{p2p} bandwidth enforcing strict design requirements when scaling the number of \glsentrylong{bs} antennas up to 100 or higher.
Based on the generic system partitioning and system requirements, a hardware platform was selected and evaluated.
It was shown that internal system configuration is within throughput and processing capabilities before the complete \glsentryshort{lumami} testbed parameters were described.
Finally, field trial results including \glsentrylong{ber} performance measurements and constellations were presented from both indoor and outdoor measurement campaigns.
The results showed that it is possible to separate up to 12 \glsentrylongpl{ue} on the same time/frequency resource when using massive MIMO.
Having established a flexible platform for testing new algorithms and digital base-band solutions we are able to take \glsentrylong{mami} from theory to real-world tests and standardization for next generation wireless systems.

\section*{Acknowledgment}
This work was funded by the Swedish foundation for strategic research SSF, VR, the strategic research area ELLIIT, and the EU Seventh Framework Programme (FP7/2007-2013) under grant agreement n 619086 (MAMMOET).

\ifCLASSOPTIONcaptionsoff
  \newpage
\fi

\bibliographystyle{MyIEEEtran}
\bibliography{references,massivemimo}
\begin{IEEEbiography}[{\includegraphics[width=1in,height=1.25in,clip,keepaspectratio]{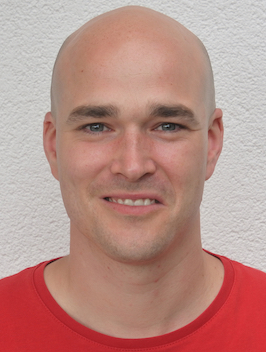}}]{Steffen Malkowsky}
	received the B.Eng. degree in Electrical Engineering and Information Technology from Pforzheim University, Germany in 2011 and the M.Sc. degree in Electronic Design from Lund University in 2013. He is currently a PhD student in the Digital ASIC group at the department of Electrical and Information Technology, Lund University. His research interests include development of reconfigurable hardware and implementation of algorithms for wireless communication with emphasis on massive MIMO. For the development of a massive MIMO testbed in collaboration with University of Bristol and National Instruments and a set spectral efficiency world record, he received 5 international awards from National Instruments, Xilinx and Hewlett Packard Enterprise.
\end{IEEEbiography}
\begin{IEEEbiography}[{\includegraphics[width=1in,height=1.25in,clip,keepaspectratio]{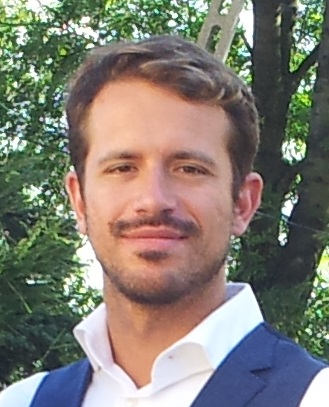}}]{João Vieira}
	received the B.Sc. degree in Electronics and Telecommunications Engineering from University of Madeira in 2011, and the M.Sc. degree in Wireless Communications from Lund University, Sweden in 2013. He is currently working towards a Ph.D. degree at the department of Electrical and Information Technology in Lund University. His main research interests regard parameter estimation and implementation issues in massive MIMO systems.
\end{IEEEbiography}
\begin{IEEEbiography}[{\includegraphics[width=1in,height=1.25in,clip,keepaspectratio]{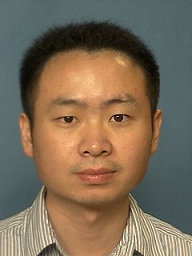}}]{Liang Liu}
	received his B.S. and Ph.D. degree in the Department of Electronics Engineering (2005) and Micro-electronics (2010) from Fudan University, China. In 2010, he was with Rensselaer Polytechnic Institute (RPI), USA as a visiting researcher. He joined Lund University as a Post-doc in 2010. Since 2016, he is Associate Professor at Lund University. His research interest includes wireless communication system and digital integrated circuits design. He is a board member of the IEEE Swedish SSC/CAS chapter. He is also a member of the technical committees of VLSI systems and applications and CAS for communications of the IEEE circuit and systems society.
\end{IEEEbiography}
\begin{IEEEbiography}[{\includegraphics[width=1in,height=1.25in,clip,keepaspectratio]{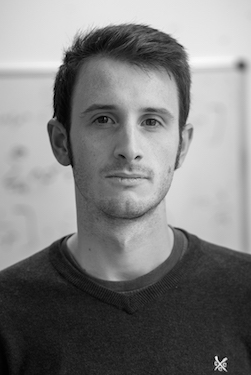}}]{Paul Harris} 
	graduated from the University of Portsmouth with a 1st Class Honours degree in Electronic Engineering in 2013 and joined the Communication Systems \& Networks Group at the University of Bristol in the same year to commence a PhD. His research interests include massive MIMO system design, performance evaluation in real-world scenarios, and the optimisation of techniques such as user grouping or power control using empirical data. Working in collaboration with Lund University and National Instruments, he implemented a 128-antenna massive MIMO test system and led two research teams to set spectral efficiency world records in 2016. For this achievement, he received 5 international awards from National Instruments, Xilinx and Hewlett Packard Enterprise, and an honorary mention in the 2016 IEEE ComSoc Student Competition for "Communications Technology Changing the World".
\end{IEEEbiography}
\begin{IEEEbiography}[{\includegraphics[width=1in,height=1.25in,clip,keepaspectratio]{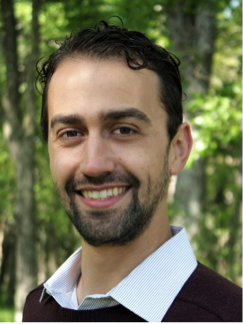}}]{Dr. Karl Nieman}
	is a Senior Wireless Platform Architect in the Advanced Wireless Research team at National Instruments.  His interests are with research and standardization of 5G technology, particularly Massive MIMO architectures and signal processing.  He has designed and implemented several FPGA-based real-time wireless communication systems, has made multiple contributions to 3GPP RAN1, and holds multiple issued and pending patents on 5G technologies.  He earned his Ph.D. and M.S. in Electrical Engineering from University of Texas at Austin in 2014 and 2011, respectively, and his B.S. in Electrical Engineering from New Mexico Institute of Mining and Technology in 2009. 
\end{IEEEbiography}
\begin{IEEEbiography}[{\includegraphics[width=1in,height=1.25in,clip,keepaspectratio]{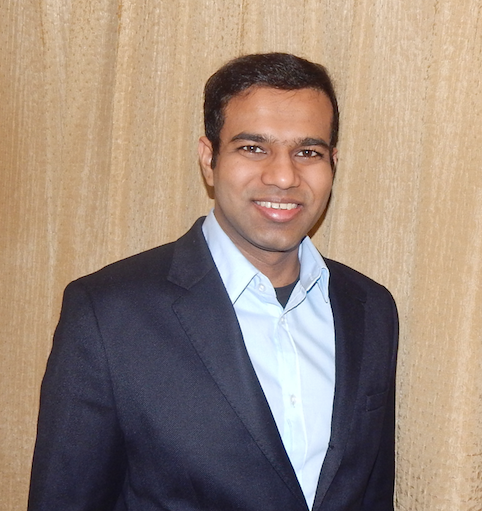}}]{Nikhil Kundargi}
	is a Senior Wireless Platform Architect in the Advanced Wireless Research Team at National Instruments since 2012. 
	He leads the Massive MIMO research initiative at NI. He is the 3GPP RAN1 Delegate for NI and participates in LTE-Advanced and 5G cellular standardization. His research interests include Massive MIMO, Full Dimension MIMO, 5G New Radio, mmWave, PHY/MAC layer design and prototyping, Dense LTE networks,  Real-time DSP, Software Defined Radio Architectures, Cognitive Radio Networks, Spectrum Sensing, Dynamic Spectrum Access, Anomaly Detection and Statistical Signal Processing. 
	He received his Ph.D in Electrical Engineering from the University of Texas at Austin in 2012. He was a member of WNCG (Wireless Networking and Communications Group at UT Austin and formerly a Graduate School Fellow at the University of Minnesota from 2006-10.  He is also an active member of IEEE Austin ComSoc Chapter and served as the Vice-Chair for the Industry Forum at IEEE Globecom 2015.
\end{IEEEbiography}
\begin{IEEEbiography}[{\includegraphics[width=1in,height=1.25in,clip,keepaspectratio]{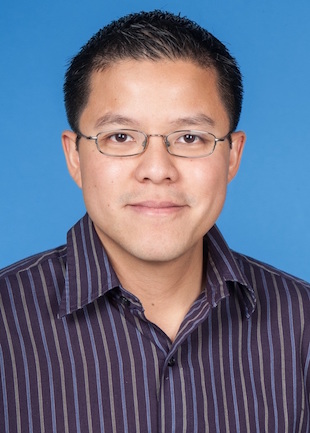}}]{Dr. Ian C. Wong}
	is Senior Manager of the Advanced Wireless Research group at National Instruments where he leads the company’s 3GPP and 802.11 wireless standards strategy and platforms for wireless system design, simulation, prototyping, and implementation.  From 2007-2009, he was a systems research and standards engineer with Freescale Semiconductor where he represented Freescale in the 3GPP LTE standardization efforts.  He was the Industry Program Chair for IEEE Globecom 2015 in Austin, the Director for Industry Communities for IEEE Communications Society 2016-2017, and a senior member of the IEEE.  His current research interests include 5G wireless systems design and prototyping, and design automation tools for rapid algorithm development. 
	Dr. Wong is the co-author of the Springer book “Resource Allocation for Multiuser Multicarrier Wireless Systems,” has over 10 patents, over 25 peer-reviewed journal and conference papers, and over 40 standards contributions. He was awarded the Texas Telecommunications Engineering Consortium Fellowship in 2003-2004, and the Wireless Networking and Communications Group student leadership award in 2007.
	He received the MS and PhD degrees in electrical engineering from the University of Texas at Austin in 2004 and 2007, respectively, and a BS degree in electronics and communications engineering (magna cum laude) from the University of the Philippines in 2000.
\end{IEEEbiography} 
\begin{IEEEbiography}[{\includegraphics[width=1in,height=1.25in,clip,keepaspectratio]{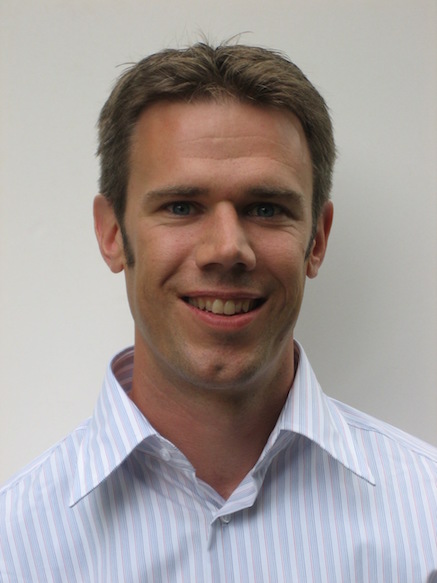}}]{Fredrik Tufvesson}
	received his Ph.D. in 2000 from Lund University in Sweden. After two years at a startup company, he joined the department of Electrical and Information Technology at Lund University, where he is now professor of radio systems. His main research interests are channel modelling, measurements and characterization for wireless communication, with applications in various areas such as massive MIMO, UWB, mm wave communication, distributed antenna systems, radio based positioning and vehicular communication. Fredrik has authored around 60 journal papers and 120 conference papers, recently he got the Neal Shepherd Memorial Award for the best propagation paper in IEEE Transactions on Vehicular Technology.
\end{IEEEbiography}
\begin{IEEEbiography}[{\includegraphics[width=1in,height=1.25in,clip,keepaspectratio]{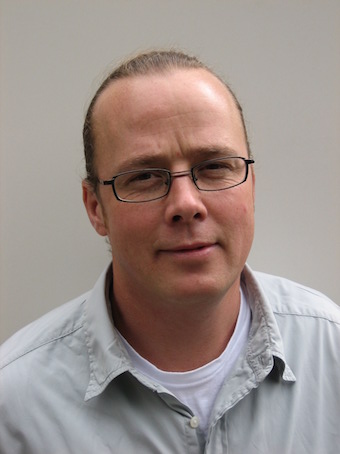}}]{Viktor \"{O}wall}
	received the M.Sc. and Ph.D. degrees in electrical engineering from Lund University, Lund, Sweden, in 1988 and 1994, respectively. 
	During 1995 to 1996, he joined the Electrical Engineering Department, the University of California at Los Angeles as a Postdoc where he mainly worked in the field of multimedia simulations. 
	Since 1996, he has been with the Department of Electrical and Information Technology, Lund University, Lund, Sweden. 
	He is currently full Professor and since 2015 the Dean of the Faculty of Engineering. 
	He was the founder and the Director of the VINNOVA Industrial Excellence Center in System Design on Silicon (SoS) which he headed until 2014. 
	His main research interest is in the field of digital hardware implementation, especially algorithms and architectures for wireless communication and biomedical applications. 
	He was co-founder of the start-up company Phase Holographic Imaging who develops microscopes utilizing digital holography.
\end{IEEEbiography}
\begin{IEEEbiography}[{\includegraphics[width=1in,height=1.25in,clip,keepaspectratio]{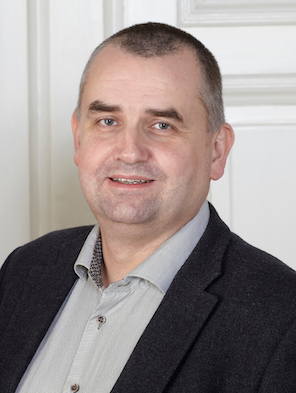}}]{Ove Edfors}
	is Professor of Radio Systems at the Department of Electrical and Information Technology, Lund University, Sweden.
	His research interests include statistical signal processing and low-complexity algorithms with applications in wireless communications. 
	In the context of Massive MIMO, his main research focus is on how realistic propagation characteristics 
	influence system performance and base-band processing complexity.
\end{IEEEbiography}
\end{document}

%% file: acronyms.tex
% Here we define our acronyms
% USAGE
% use \newacronym{label}{short}{long} for new entries
% call the entries with following command
% \gls{label} to call
% \glspl{lable} to call plural
% \Gls{label} to call with first letter of long name capitalized
% \Glspl{label} to call with first letter of long name capitalized and plural
% If you have an entry where you need a special pluralization (instead of just adding an s to the last word) use the longplural option
% \newacronym[longplural=xxx]{label}{short}{long}

% How to compile
% run pdflatex and afterwards makeglossaries and then pdflatex again

\newacronym[longplural=downlink data]{dld}{DLD}{downlink data}
\newacronym[longplural=uplink data]{uld}{ULD}{uplink data}
\newacronym{adc}{ADC}{analog-to-digital converter}
\newacronym{afe}{AFE}{analog-front end}
\newacronym{agu}{AGU}{address generation unit}
\newacronym{ahb}{AHB}{Advanced High-Performance Bus}
\newacronym{asip}{ASIP}{application-specific instruction processor}
\newacronym{awgn}{AWGN}{additive white Gaussian noise}
\newacronym{bb}{BB}{baseband}
\newacronym{ber}{BER}{Bit Error Rate}
\newacronym{bi}{BI}{bus interface}
\newacronym{bs}{BS}{base station}
\newacronym{bw}{BW}{bandwidth}
\newacronym{ce}{CE}{channel estimation}
\newacronym{cm}{CM}{core manager}
\newacronym{cp}{CP}{cyclic prefix}
\newacronym{cpu}{CPU}{central processing unit}
\newacronym{csi}{CSI}{channel state information}
\newacronym{cs}{CS}{circuit-switched}
\newacronym{dac}{DAC}{digital-to-analog converter}
\newacronym{dfe}{DFE}{digital front end}
\newacronym{dlp}{DLP}{data-level parallelism}
\newacronym{dl}{DL}{down-link}
\newacronym{dma}{DMA}{direct memory access}
\newacronym{dsp}{DSP}{digital signal processing}
\newacronym{dram}{DRAM}{Dynamic \gls{ram}}
\newacronym{dvs}{DVS}{dynamic voltage scaling}
\newacronym{ecc}{ECC}{error-correcting code}
\newacronym{eda}{EDA}{electronic design automation}
\newacronym{evm}{EVM}{error vector magnitude}
\newacronym{fdd}{FDD}{frequency-division duplex}
\newacronym{fe}{FE}{Front-End}
\newacronym{fft}{FFT}{fast-Fourier transform}
\newacronym{fifo}{FIFO}{first-in first-out}
\newacronym{fpga}{FPGA}{Field Programmable Gate Array}
\newacronym{gals}{GALS}{globally-asynchronous locally-synchronous}
\newacronym{ggm}{GGM}{general Gram matrix}
\newacronym{gpp}{GPP}{General Purpose Processor}
\newacronym{gpu}{GPU}{graphical processing unit}
\newacronym{gp}{GP}{general-purpose}
\newacronym{gps}{GPS}{Global Positioning System}
\newacronym{hdl}{HDL}{hardware description language}
\newacronym{hls}{HLS}{high-level synthesis}
\newacronym{hw}{HW}{hardware}
\newacronym{ifft}{IFFT}{inverse \gls{fft}}
\newacronym{iid}{iid}{independent and identically distributed}
\newacronym{ip}{IP}{intellectual property}
\newacronym{iot}{IoT}{internet of things}
\newacronym{ltea}{LTE-A}{long-term evolution advanced}
\newacronym{lte}{LTE}{long-term evolution}
\newacronym{lumami}{LuMaMi}{Lund University \gls{mami}}
\newacronym{lut}{LUT}{lookup-table}
\newacronym{macl}{MACL}{medium-access control layer}
\newacronym{mac}{MAC}{multiply-accumulate units}
\newacronym{mami}{MaMi}{massive \gls{mimo}}
\newacronym{mf}{MF}{matched-filter}
\newacronym{mimo}{MIMO}{multiple-input multiple-output}
\newacronym{mrc}{MRC}{maximum ratio combining}
\newacronym{mrt}{MRT}{maximum ratio transmission}
\newacronym{mumimo}{MU-MIMO}{multi-user \gls{mimo}}
\newacronym{mmse}{MMSE}{minimum \gls{mse}}
\newacronym{mse}{MSE}{mean-square error}
\newacronym{ni}{NI}{National Instruments}
\newacronym{noc}{NoC}{network-on chip}
\newacronym{ocp}{OCP}{Open Core Protocol}
\newacronym{ofdm}{OFDM}{Orthogonal Frequency Division Multiplexing}
\newacronym{pa}{PA}{power amplifier}
\newacronym{p2p}{P2P}{peer-to-peer}
\newacronym{pap}{PAP}{per-antenna processing}
\newacronym{pcie}{PCIe}{Peripheral Component Interconnect Express}
\newacronym{pe}{PE}{processing element}
\newacronym{per}{PER}{packet error rate}
\newacronym{psp}{PSP}{per-subcarrier processing}
\newacronym{ps}{PS}{packet-switched}
\newacronym{pup}{PUP}{per-user processing}
\newacronym{papr}{PAPR}{peak-to-average power ratio}
\newacronym{pss}{PSS}{Primary Synchronisation Signal}
\newacronym{qam}{QAM}{quadrature amplitude modulation}
\newacronym{ram}{RAM}{Random Access Memory}
\newacronym{rf}{RF}{radio-frequency}
\newacronym{rlc}{RLC}{Reconfigurable Logic Core}
\newacronym{rgf}{RGF}{register-file}
\newacronym{risc}{RISC}{reduced-instruction set computer}
\newacronym{rx}{RX}{receiver}
\newacronym{rzf}{RZF}{regularized \gls{zf}}
\newacronym{sdr}{SDR}{Software-Defined Radio}
\newacronym{simd}{SIMD}{single instruction multiple data}
\newacronym{snr}{SNR}{signal-to-noise ratio}
\newacronym{soc}{SoC}{system-on chip}
\newacronym{tdd}{TDD}{time-division duplex}
\newacronym{tdm}{TDM}{time-division multiplexing}
\newacronym{tx}{TX}{transmitter}
\newacronym{ota}{OTA}{over-the-air}
\newacronym{ue}{UE}{user equipment}
\newacronym{upa}{UPA}{user processing accelerator}
\newacronym{ulp}{ULP}{uplink pilot}
\newacronym{ul}{UL}{up-link}
\newacronym{vliw}{VLIW}{very large instruction word}
\newacronym{vlsi}{VLSI}{very-large scale integration}
\newacronym{zf}{ZF}{zero-forcing}
\newacronym{plm}{PLM}{parallel memory}

%% file: commands.tex
\newcommand{\ie}{\emph{i.e., }}
\newcommand{\eg}{\emph{e.g., }}

 \newcounter{comment}

\newcommand*\circled[1]{\tikz[baseline=(char.base)]{
\node[shape=circle,draw,inner sep=2pt] (char) {#1};}}

\DeclarePairedDelimiter\ceil{\lceil}{\rceil}

%% file: subsystem.tikz
%\resizebox{\columnwidth}{!}{
\begin{tikzpicture}
\linespread{0.7}
% include the figure
\node[anchor=south west, inner sep=0pt] at (0,0) {\includegraphics[width=0.9\columnwidth]{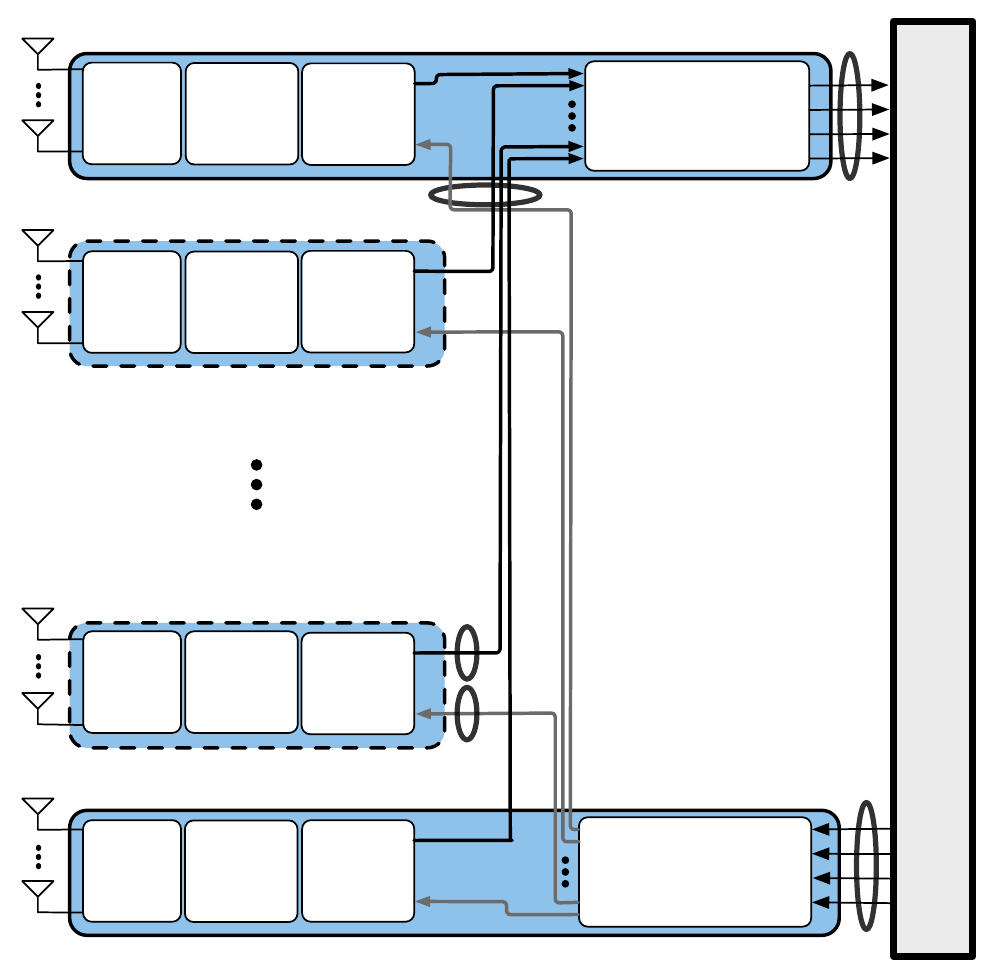}};
% Draw grid to place text easier
%\draw[step=.5cm,very thin,dashed,color=black] (0,0) grid (8,8);
%\draw[step=1cm,very thin,dashed,color=red] (0,0) grid (8,8);
%\draw[step=5cm,very thin,dashed,color=blue] (0,0) grid (8,8);
%\fill (0,0) circle (2pt);
% Router text
\node[align=center,anchor=south west] at (4.75,6.55) {\scriptsize $n_\text{sub}n_\text{ant}:n_\text{co}$\\\scriptsize Router};
\node[align=center,anchor=south west] at (4.75,0.5) {\scriptsize $n_\text{co}:n_\text{sub}n_\text{ant}$\\\scriptsize Router};
% RF Front ends
\node[align=center,anchor=south west] (RF1) at (0.675,0.5) {\scriptsize RF\\\scriptsize Front};
\node[align=center,above=0.8cm of RF1.north,anchor=south] (RF2) {\scriptsize RF\\\scriptsize Front};
\node[align=center,above=2.35cm of RF2.north,anchor=south] (RF3) {\scriptsize RF\\\scriptsize Front};
\node[align=center,above=0.8cm of RF3.north,anchor=south] (RF4) {\scriptsize RF\\\scriptsize Front};
% Reci. Comp.
\node[align=center,right=0.06cm of RF1.east,anchor=west] (RC1) {\scriptsize Reci.\\\scriptsize Comp.};
\node[align=center,right=0.06cm of RF2.east,anchor=west] (RC2) {\scriptsize Reci.\\\scriptsize Comp.};
\node[align=center,right=0.06cm of RF3.east,anchor=west] (RC3) {\scriptsize Reci.\\\scriptsize Comp.};
\node[align=center,right=0.06cm of RF4.east,anchor=west] (RC4) {\scriptsize Reci.\\\scriptsize Comp.};
% OFDM
\node[align=center,right=-0.025cm of RC1.east,anchor=west] (O1) {\scriptsize OFDM\\\scriptsize TX/RX};
\node[align=center,right=-0.025cm of RC2.east,anchor=west] (O2) {\scriptsize OFDM\\\scriptsize TX/RX};
\node[align=center,right=-0.025cm of RC3.east,anchor=west] (O3) {\scriptsize OFDM\\\scriptsize TX/RX};
\node[align=center,right=-0.025cm of RC4.east,anchor=west] (O4) {\scriptsize OFDM\\\scriptsize TX/RX};
% BUS
\node[align=center,anchor=north] (bus1) at (7.5,4.8) {\large B};
\node[align=center,below=-0.1cm of bus1.south,anchor=north] (bus2) {\large U};
\node[align=center,below=-0.14cm of bus2.south,anchor=north] (bus3) {\large S};
% Please numbering for rates
\node[anchor=south west] at (4.4,5.8) {\circled{\footnotesize 1}};
\node[anchor=south west] at (6.4,7.4) {\circled{\footnotesize 1}};
\node[anchor=south west] at (6.5,1.35) {\circled{\footnotesize 1}};
\node[anchor=south west] at (3.9,2.05) {\circled{\footnotesize 2}};
% Place numbering for USRPs
\node[anchor=south west] at (0.6,7.4) {\footnotesize $1$};
\node[anchor=south west] at (0.6,5.9) {\footnotesize $2$};
\node[anchor=south west] at (0.6,2.85) {\footnotesize $n_\text{sub}-1$};
\node[anchor=south west] at (0.6,1.3) {\footnotesize $n_\text{sub}$};
% Please legend
\node[anchor= south west,draw,align=left] at (1.5,-1) {\footnotesize\circled{1} $R_{\text{SDR}_\text{in/out}}=n_\text{ant}n_\text{sub}wF_\text{s}\,\SI{}{MB/s}$\\\footnotesize\circled{2} $(n_\text{ant})(wF_\text{s})\,\SI{}{MB/s}$};
\end{tikzpicture}
%}

%% file: co_processor.tikz
%\resizebox{\columnwidth}{!}{
\begin{tikzpicture}
\linespread{0.7}
% insert figure
\node[anchor=south west, inner sep=0pt] at (0,0) {\includegraphics[width=1\columnwidth]{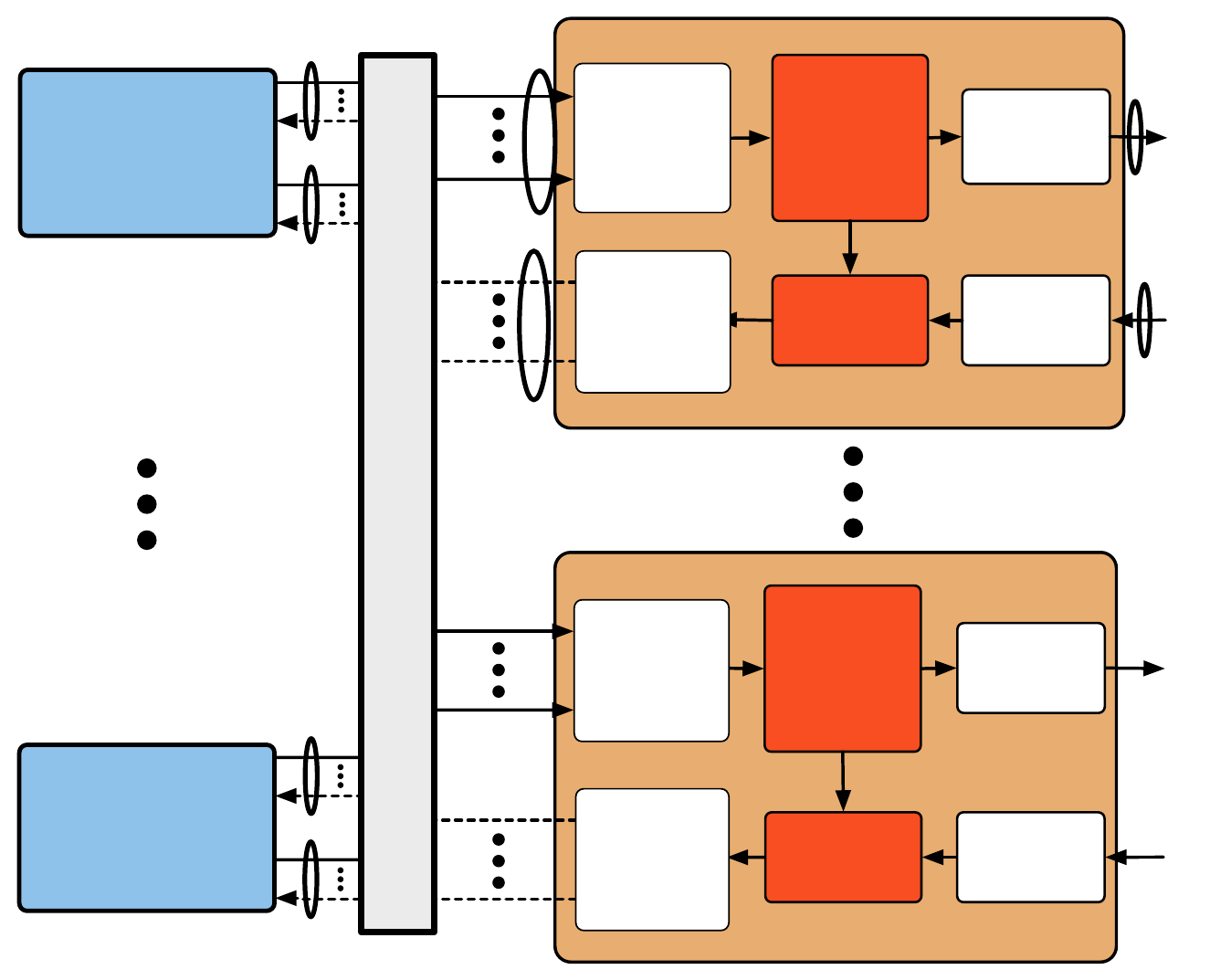}};
% draw grid to place text easier
%\draw[step=.5cm,very thin,dashed,color=black] (0,0) grid (8,8);
%\draw[step=1cm,very thin,dashed,color=red] (0,0) grid (8,8);
%\draw[step=5cm,very thin,dashed,color=blue] (0,0) grid (8,8);
%\fill (0,0) circle (2pt);
% CE and detection
\node[align=center,text width=1cm,anchor=south west] (CE1) at (5.55,1.62) {\scriptsize Channel Esti. + MIMO Detection};
\node[align=center,text width=1cm,anchor=south west] (CE2) at (5.6,5.5) {\scriptsize Channel Esti. + MIMO Detection};
% Precoding
\node[align=center,below=.35cm of CE1.south,text width=1cm,anchor=north] (Pre1) {\scriptsize MIMO Precoding};
\node[align=center,below=.3cm of CE2.south,text width=1cm,anchor=north] (Pre2) {\scriptsize MIMO Precoding};
% Symbol Demap/Map
\node[align=center,text width=1cm,anchor=south west] (DE1) at (6.95,1.875) {\scriptsize Symbol Demap};
\node[align=center,below=0.65cm of DE1.south,text width=1cm,anchor=north] (M1) {\scriptsize Symbol Map};
\node[align=center,text width=1cm,anchor=south west] (DE2) at (6.95,5.775) {\scriptsize Symbol Demap};
\node[align=center,below=0.6cm of DE2.south,text width=1cm,anchor=north] (M2) {\scriptsize Symbol Map};
% Routers Typ A
\node[align=center,left=0.17cm of CE1.west,text width=1cm,anchor=east] (R1A) {\scriptsize Router Typ A};
\node[align=center,left=0.2cm of CE2.west,text width=1cm,anchor=east] (R1A) {\scriptsize Router Typ A};
% Routers Typ B
\node[align=center,left=0.15cm of Pre1.west,text width=1cm,anchor=east] (R1B) {\scriptsize Router Typ B};
\node[align=center,left=0.2cm of Pre2.west,text width=1cm,anchor=east] (R1B) {\scriptsize Router Typ B};
% Co processor numbering
\node[align=center,text width=1cm,anchor=south west] (CO1) at (7.25,7.1) {\footnotesize 1};
\node[align=center,below=3.6cm of CO1.south,anchor=north] (COnco) {\footnotesize $n_\text{co}$};
% Data rate numbering
\node[anchor=south west] (num2_1) at (8.25,6.2) {\circled{\footnotesize 2}};
\node[align=center,below=1.52cm of num2_1.south,anchor=north] (num2_2) {\circled{\footnotesize 2}};
\node[anchor=south west] (num1_1) at (3.45,6.6) {\circled{\footnotesize 1}};
\node[align=center,below=2.25cm of num1_1.south,anchor=north] (num1_2) {\circled{\footnotesize 1}};
% antenna streams
\node[anchor=south east] (a1) at (3.73,5.87) {\circled{\scriptsize a}};
\node[align=center,below=.72cm of a1.south,anchor=north] (a2) {\circled{\scriptsize a}};
\node[align=center,below=1.96cm of a2.south,anchor=north] (a3) {\circled{\scriptsize a}};
\node[align=center,below=0.78cm of a3.south,anchor=north] (a4) {\circled{\scriptsize a}};
% subband markers
\node[anchor=south east] (sub1) at (2.6,6.7) {sub-band 1};
\node[align=center,below=1.25cm of sub1.south,anchor=north] (sub2) {sub-band $n_\text{co}$};
\node[align=center,below=2.7cm of sub2.south,anchor=north] (sub3) {sub-band 1};
\node[align=center,below=1.25cm of sub3.south,anchor=north] (sub4) {sub-band $n_\text{co}$};
% BUS
\node[align=center,anchor=north] (bus1) at (2.92,4.5) {\large B};
\node[align=center,below=-0.1cm of bus1.south,anchor=north] (bus2) {\large U};
\node[align=center,below=-0.14cm of bus2.south,anchor=north] (bus3) {\large S};
% Router Types Legend
\node[anchor= south west,draw,align=left] at (5,-2) {Router Types\\\footnotesize Typ A: $\left(\left\lceil\frac{M}{n_\text{sub}n_\text{ant}}\right\rceil\right) : (n_\text{sub})$\\\footnotesize Typ B: $(n_\text{sub}) : \left(\left\lceil\frac{M}{n_\text{sub}n_\text{ant}}\right\rceil\right)$};
% Legend
\node[anchor= south west,draw,align=left] at (-0.2,-2) {\footnotesize\circled{a} $\text{\# antenna streams / link}=n_\text{sub}n_\text{ant}$\\\footnotesize\circled{1} $R_{\text{CO}_\text{in/out}}=M/n_\text{co}wF_\text{s}\,\SI{}{MB/s}$\\\footnotesize\circled{2} $K\,\SI{16.8}{\mega B\per\second}$};
% Subsytems
\linespread{1}
\node[align=center,anchor=south west] (Subsys1) at (0.195,5.6) {Subsystem\\1};
\node[align=center,below=4cm of Subsys1.south,anchor=north] {Subsystem\\\footnotesize$\ceil*{M/(n_\text{sub}n_\text{ant})}$};
\end{tikzpicture}
%}

%% file: Grouped_BERs.tikz
 	\begin{tikzpicture}
 	\begin{groupplot}[
	 	group style={group name=my plots,group size= 2 by 2,%
	 		vertical sep=1.2cm, horizontal sep = 0.5cm,%
	 		xticklabels at=edge bottom, yticklabels at=edge left},
	 	ymin=1e-4,
	 	ymax=1e0,
	 	xmin=0,
	 	xmax=30,
	 	xtick={0,5,...,30},
	 	minor x tick num=4,
	 	xmajorgrids, 
	 	ymajorgrids,	 	
	 	cycle list name=my cycle list,
	 	%smooth,
	 	ytick align=outside,
	 	xtick align=outside,
	 	width=0.8\columnwidth, height=0.19\paperheight]
 	\nextgroupplot[mark size=1.2, ymode=log, ylabel={Uncoded BER}]
	 	\foreach \yindex in {1,...,12}
	 	{   
	 		\addplot table [y index = \yindex] {UL_all_users_qpsk.dat};\label{plots:plot\yindex}
	 	} 
	 	\coordinate (top) at (rel axis cs:0,1);% coordinate at top of the first plot
 	\nextgroupplot[mark size=1.2, ymode=log]
	 	\foreach \yindex in {1,...,12}
	 	{   
	 		\addplot table [y index = \yindex] {UL_all_users_64.dat};
	 	} 
 	\nextgroupplot[mark size=1.2, ymode=log, ylabel={Uncoded BER}, xlabel={Amplifier Gain in dB}]
	 	\foreach \yindex in {1,...,12}
	 	{   
	 		\addplot table [y index = \yindex] {DL_all_users_qpsk.dat};
	 	} 
 	\nextgroupplot[mark size=1.2, ymode=log, xlabel={Amplifier Gain in dB}]
	 	\foreach \yindex in {1,...,12}
	 	{   
	 		\addplot table [y index = \yindex] {DL_all_users_64.dat};
	 	} 
 	\coordinate (bot) at (rel axis cs:1,0);% coordinate at bottom of the last plot
 	\end{groupplot}
 	\node[below = 0.4cm of my plots c1r1.south] {(a) UL with QPSK};
 	\node[below = 0.4cm of my plots c2r1.south] {(b) UL with 64-QAM};
 	\node[below = 1.2cm of my plots c1r2.south] {(c) DL with QPSK};
 	\node[below = 1.2cm of my plots c2r2.south] {(d) DL with 64-QAM};
 	%\node[text width=6cm,align=center,anchor=north] at ([yshift=-5mm]my plots c2r1.south) {\captionof{subfigure}{Pressure relative (P/P)\label{subplot:two}}};
% Uncomment if text on the left for all y-axis needed
% 	\path (top-|current bounding box.west)--
% 	node[anchor=south,rotate=90] {TEXT GOES HERE}
% 	(bot-|current bounding box.west);
 	% legend
 	\path (top|-current bounding box.north)--
 	coordinate(legendpos)
 	(bot|-current bounding box.north);
 	\matrix[
 	matrix of nodes,
 	anchor=south,
 	draw,
 	inner sep=0.2em,
 	draw
 	]at([yshift=1ex]legendpos)
 	{
	 	% foreach loop unfortunately not working here...
 		\ref{plots:plot1}& UE0&[5pt]
 		\ref{plots:plot2}& UE1&[5pt]
 		\ref{plots:plot3}& UE2&[5pt]
 		\ref{plots:plot4}& UE3&[5pt]
 		\ref{plots:plot5}& UE4&[5pt]
 		\ref{plots:plot6}& UE5&[5pt]\\
 		\ref{plots:plot7}& UE6&[5pt]
 		\ref{plots:plot8}& UE7&[5pt]
 		\ref{plots:plot9}& UE8&[5pt]
 		\ref{plots:plot10}& UE9&[5pt]
 		\ref{plots:plot11}& UE10&[5pt]
 		\ref{plots:plot12}& UE11\\};
 	\end{tikzpicture}

%% file: BER_plot_AWGN.tikz
% Cycle list for pgf plots
\pgfplotscreateplotcyclelist{my cycle list}{%
	solid, black, every mark/.append style={solid}, mark=*\\%
	solid, red, every mark/.append style={solid}, mark=square*\\%
	solid, blue, every mark/.append style={solid}, mark=triangle*\\%
}

	\begin{tikzpicture}[every node/.style={font=\footnotesize}]
	\begin{semilogyaxis}[
				ymode=log,
				ylabel={Uncoded BER},
				ylabel style ={yshift = -0.10cm},
				xlabel={SNR (in dB)},
				xlabel style ={yshift = +0.15cm},
				width=.85\columnwidth, height=0.2\paperheight,
				ymin=1e-4,
				ymax=1e0,
				xmin=-7,
				xmax=23,
				xtick={-5,5,...,20},
				minor x tick num=4,
				xmajorgrids, ymajorgrids,
				legend entries={UE4,AWGN,Rayleigh},
                table/x index={0},
                mark size=1.2,
                smooth,
				legend style={nodes={scale=0.75, transform shape},%
				at={(1.2,0.0)},anchor=south east, font=\scriptsize},
				cycle list name=my cycle list]

			\foreach \yindex in {1,...,3}
			{   
				\addplot table [y index = \yindex] {\filename};
			} 	
			
%			\addplot [solid, black,no markers, line width = 1.5] table [y index = 9] {./figures/tikz/data/UL_all_users_qpsk_with_awgn.dat};
%			\addplot [dashed, black,no markers, line width = 1.5] table [y index = 10] {./figures/tikz/data/UL_all_users_qpsk_with_awgn.dat};
							
		\end{semilogyaxis}
%		\node[draw,align=right]  {some text\\ spanning three lines\\  manual line breaks};
%\matrix[ampersand replacement=\&] at (4.9,3.3) {
%	\node (dsymdsc) [fill=white,font=\scriptsize] {
%		\begin{tabular}{| c| c| c|}
%		\hline
%		Mode & D$_{sym}$ &  D$_{sc}$ \\
%		\hline
%		TD & $12$ & $1$\\
%		\hline
%		FD-TD & $12$ & $12$, $8$\\
%		\hline
%		\end{tabular}
%	};
%	\& 
%\\
%};

\end{tikzpicture}

%% file: MRC_vs_ZF.tikz
% Cycle list for pgf plots
\pgfplotscreateplotcyclelist{my cycle list}{%
	solid, black, every mark/.append style={solid}, mark=*\\%
	solid, red, every mark/.append style={solid}, mark=square*\\%
	solid, blue, every mark/.append style={solid}, mark=triangle*\\%
	densely dotted, black, every mark/.append style={solid},mark=*\\%
	densely dotted, red, every mark/.append style={solid},mark=square*\\%
	densely dotted, blue, every mark/.append style={solid},mark=triangle*\\%
}

	\begin{tikzpicture}[every node/.style={font=\footnotesize}]
	\begin{semilogyaxis}[
				ymode=log,
				ylabel={Uncoded BER},
				xlabel={Amplifier Gain (in dB)},
				xlabel style ={yshift = +0.15cm},
				width=.8\columnwidth, height=0.19\paperheight,
				ymin=1e-4,
				ymax=1e0,
				xmin=0,
				xmax=30,
				xtick={0,5,...,30},
				minor x tick num=4,
				xmajorgrids, ymajorgrids,
				legend entries={\detused\ QPSK, \detused\ 16-QAM, \detused\ 64-QAM,%
					ZF QPSK, ZF 16-QAM, ZF 64-QAM},
                table/x index={0},
                mark size=1,
                smooth,
                legend cell align=left,
				legend style={align=left, nodes={scale=0.65, transform shape},%
				at={(1.2,0.0)},anchor=south east, font=\scriptsize},
				cycle list name=my cycle list]

			\foreach \yindex in {1,...,6}
			{   
				\addplot table [y index = \yindex] {\filename};
			} 	
							
		\end{semilogyaxis}
%		\node[draw,align=right]  {some text\\ spanning three lines\\  manual line breaks};
%\matrix[ampersand replacement=\&] at (4.9,3.3) {
%	\node (dsymdsc) [fill=white,font=\scriptsize] {
%		\begin{tabular}{| c| c| c|}
%		\hline
%		Mode & D$_{sym}$ &  D$_{sc}$ \\
%		\hline
%		TD & $12$ & $1$\\
%		\hline
%		FD-TD & $12$ & $12$, $8$\\
%		\hline
%		\end{tabular}
%	};
%	\& 
%\\
%};

\end{tikzpicture}